\documentclass[twoside,twocolumn,9pt]{article}

\usepackage[colorlinks=true, linkcolor=black, citecolor=black, urlcolor=black]{hyperref}
\usepackage{xr}

\usepackage{extsizes}
\usepackage[super,sort&compress,comma]{natbib} 
\usepackage[version=3]{mhchem}
\usepackage[left=1.5cm, right=1.5cm, top=1.785cm, bottom=2.0cm]{geometry}
\usepackage{balance}
\usepackage{mathptmx}
\usepackage{sectsty}
\usepackage{graphicx} 
\usepackage{lastpage}
\usepackage[format=plain,justification=justified,singlelinecheck=false,font={stretch=1.125,small,sf},labelfont=bf,labelsep=space]{caption}
\usepackage{float}
\usepackage{fancyhdr}
\usepackage{fnpos}
\usepackage{bm}
\usepackage[english]{babel}
\addto{\captionsenglish}{%
  
}
\usepackage{array}
\usepackage{droidsans}
\usepackage{charter}
\usepackage[T1]{fontenc}
\usepackage[usenames,dvipsnames]{xcolor}
\usepackage{setspace}
\usepackage[compact]{titlesec}
\usepackage{hyperref}
\usepackage{subcaption}

\usepackage{epstopdf}

\definecolor{cream}{RGB}{222,217,201}
\begin{document}

\pagestyle{fancy}
\thispagestyle{plain}
\fancypagestyle{plain}{

\renewcommand{\headrulewidth}{0pt}
}

\makeFNbottom
\makeatletter
\renewcommand\LARGE{\@setfontsize\LARGE{15pt}{17}}
\renewcommand\Large{\@setfontsize\Large{12pt}{14}}
\renewcommand\large{\@setfontsize\large{10pt}{12}}
\renewcommand\footnotesize{\@setfontsize\footnotesize{7pt}{10}}
\makeatother

\renewcommand{\thefootnote}{\fnsymbol{footnote}}
\renewcommand\footnoterule{\vspace*{1pt}%
\color{cream}\hrule width 3.5in height 0.4pt \color{black}\vspace*{5pt}} 
\setcounter{secnumdepth}{5}

\makeatletter 
\renewcommand\@biblabel[1]{#1}            
\renewcommand\@makefntext[1]%
{\noindent\makebox[0pt][r]{\@thefnmark\,}#1}
\makeatother 
\renewcommand{\figurename}{\small{Fig.}~}
\sectionfont{\sffamily\Large}
\subsectionfont{\normalsize}
\subsubsectionfont{\bf}
\setstretch{1.125}
\setlength{\skip\footins}{0.8cm}
\setlength{\footnotesep}{0.25cm}
\setlength{\jot}{10pt}
\titlespacing*{\section}{0pt}{4pt}{4pt}
\titlespacing*{\subsection}{0pt}{15pt}{1pt}

\fancyfoot{}
\fancyfoot[LO,RE]{\vspace{-7.1pt}} 
\fancyfoot[CO]{\vspace{-7.1pt}\hspace{13.2cm}} 
\fancyfoot[CE]{\vspace{-7.2pt}\hspace{-14.2cm}} 
\fancyfoot[LE]{\footnotesize{\sffamily{\thepage~\textbar\hspace{3.45cm} 1--\pageref{LastPage}}}}
\fancyhead{}
\renewcommand{\headrulewidth}{0pt} 
\renewcommand{\footrulewidth}{0pt}
\setlength{\arrayrulewidth}{1pt}
\setlength{\columnsep}{6.5mm}
\setlength\bibsep{1pt}

\makeatletter 
\newlength{\figrulesep} 
\setlength{\figrulesep}{0.5\textfloatsep} 

\newcommand{\topfigrule}{\vspace*{-1pt}%
\noindent{\color{cream}\rule[-\figrulesep]{\columnwidth}{1.5pt}} }

\newcommand{\botfigrule}{\vspace*{-2pt}%
\noindent{\color{cream}\rule[\figrulesep]{\columnwidth}{1.5pt}} }

\newcommand{\dblfigrule}{\vspace*{-1pt}%
\noindent{\color{cream}\rule[-\figrulesep]{\textwidth}{1.5pt}} }

\makeatother

\twocolumn[
  \begin{@twocolumnfalse}
{\hfill\raisebox{0pt}[0pt][0pt]{}\\[1ex]
}\par
\vspace{1em}
\sffamily
\begin{tabular}{m{1.5cm} p{15cm} }

 & \noindent\LARGE{\textbf{Effective interactions in active Brownian particles$^\dag$}} \\
\vspace{0.3cm} & \vspace{0.3cm} \\

 & \noindent\large{Clare R. Rees-Zimmerman,$^{\ast}$\textit{$^{a}$} C. Miguel Barriuso Gutierrez,\textit{$^{b\ddag}$} Chantal Valeriani\textit{$^{b\ddag}$} and Dirk G. A. L. Aarts\textit{$^{a}$}} \\

 & \noindent\normalsize{We report an approach to obtain effective pair potentials which describe the structure of two-dimensional systems of active Brownian particles. The pair potential is found by an inverse method, which matches the radial distribution function found from two different schemes. The inverse method, previously demonstrated via simulated equilibrium configurations of passive particles, has now been applied  to a suspension of active particles. Interestingly, although active particles are inherently not in equilibrium, we still obtain effective interaction potentials which accurately describe the structure of the active system. Treating these effective potentials as if they were those of equilibrium systems, furthermore allows us to measure effective chemical potentials and pressures. Both the passive interactions and active motion of the active Brownian particles contribute to their effective interaction potentials.} \\

\end{tabular}

 \end{@twocolumnfalse} \vspace{0.6cm}

  ]

\renewcommand*\rmdefault{bch}\normalfont\upshape
\rmfamily
\section*{}
\vspace{-1cm}

\footnotetext{\textit{$^{a}$~Physical and Theoretical Chemistry Laboratory, University of Oxford, South Parks Road, Oxford OX1 3QZ, United Kingdom. Email: clare.rees-zimmerman@chch.ox.ac.uk or dirk.aarts@chem.ox.ac.uk}}
\footnotetext{\textit{$^{b}$~Departamento de Estructura de la Materia, Física Térmica y Electrónica, Universidad Complutense de Madrid, 28040 Madrid, Spain. Email: carbarri@ucm.es or cvaleriani@ucm.es}} 
\footnotetext{\dag~Supplementary Information available. See DOI: 00.0000/xxxxxxxxxx/}

\footnotetext{\ddag~Also at Grupo Interdisciplinar Sistemas Complejos, Madrid, Spain}

\section{Introduction}\label{sec:Intro}
Dispersions of active particles are non-equilibrium systems: they dissipate energy in powering an active component to the particle velocities.\cite{khadka2018,smith2023} The collective behaviour of active particles manifests both structural and dynamic effects, as observed in, e.g., flocks of birds, schools of fish, and suspensions of bacteria.\cite{negi2022} Inspired by natural systems, synthetic active particle systems can be designed to follow a particular behaviour, through tuning the chemistry or physics of colloidal particles and/or their external fields.\cite{shields2017} Modelling active particles as active \textit{Brownian} particles (ABPs) provides a simple means to better understand their behaviour; these are Brownian particles with an added self-propulsion, the direction of which diffuses over time.\cite{zottl2023, stenhammar2014}

Structural effects observed in suspensions of active particles include flocking,\cite{caprini2023} clustering,\cite{navarro2015, mognetti2013} chain and ring formation,\cite{telezki2020} and motility-induced phase separation (MIPS), which occurs in active Brownian particles due to particles slowing down in higher density regions.\cite{cates2015, stenhammar2014,mallory2018,martinroca2021} The physics of active Brownian particles comprises passive interactions and the particles’ active motion.\cite{mauleonamieva2020} So far, it has  not proved straightforward to find a thermodynamic framework for active matter. For example, it appears that mechanical pressure is only a state variable for active particles in certain systems.\cite{junot2017,solon2015,martinroca2021}

\textcolor{black}{In the framework for ABPs, the passive interactions are described by the pair potential $u(r)$, while the active component is controlled by parameters such as the self-propulsion force and translational and rotational diffusion coefficients. A convenient dimensionless measure of activity is the P\'eclet number, which compares propulsion to diffusive motion.\cite{stenhammar2013} The pair potential $u(r)$ specifies the direct interaction between two particles at separation
$r$. Together with temperature and density, depending on the form of $u(r)$ --- e.g, hard-core repulsion,\cite{percus1958} soft repulsion\cite{stillinger1978} or Lennard-Jones type interaction\cite{lennardjones1931} --- it sets the equilibrium structure that would be obtained in the absence of activity.} 

\textcolor{black}{A steady radial distribution function, $g(r)$, can be obtained from ABPs given the particle number density, pair interaction, $u(r)$, and the strength of self-propulsion relative to diffusion. In this way, ABPs can form a steady state structure with a reproducible $g(r)$, in a way similar to equilibrium systems.\cite{evans2024}} In this work\textcolor{black}{,} we aim to address whether an equilibrium approach could help us to understand the non-equilibrium structure of active particles. \textcolor{black}{We seek to do this by finding effective pair potentials of an equivalent passive system. We begin by noting that a variety of definitions for effective potentials have been defined in previous works, alongside the warning of \citet{louis2002}, that --- in systems with density-dependent interactions --- there is no single effective pair potential; the correct effective potential needs to be derived according to its desired use.} 

\textcolor{black}{Perhaps the simplest definition of an effective pair potential is the potential of mean force, $w(r)$.\cite{kirkwood1935} Physically, the potential of mean force, 
$w(r)$, represents the effective interaction felt between two particles due to both their direct pair interaction and the averaged influence of all other particles in the system. It is related to the radial distribution function through $g(r)=-\exp[\beta w(r)]$. However, only in the dilute limit does an equilibrium simulation using $w(r)$ fully reproduce the observed $g(r)$.} Microscopy experiments on active particles have been conducted with very dilute dispersions, \textcolor{black}{in which case the potential of mean force is an appropriate choice of effective pair potential.\cite{mu2022}} 

\textcolor{black}{Our approach seeks effective pair potentials which, when simulated as a passive system (with only these effective pairwise interactions), can regenerate the structure of the ABPs, as described by a $g(r)$ --- and not just in the dilute limit. This contrasts with, for example, the approach of \citet{turci2021}, who also aimed to recreate the structure of the active systems. They did this by separately finding 2,3,...$N$-body interactions of ABPs by simulating systems containing 2,3,...$N$ particles, respectively. Their effective $N$-body interactions then do not depend on density. Our approach instead finds only an effective pair potential $u_{\rm{eff}}(r)$ which alone regenerates $g(r)$. Since ABPs do have effective higher-order interactions, this $u_{\rm{eff}}(r)$ accounts for the effect of these. We will therefore find our $u_{\rm{eff}}(r)$ to be density-dependent. In this text, $u_{\rm{eff}}(r)$ is used to refer specifically to our definition of effective pair potential.} 

Due to the complexity of the relationship between $g(r)$ and $u(r)$ for a passive system –-- related by a simulation, or approximated using perturbation theory\cite{gottschalk2021} –-- we do not expect there to be an analytical relationship between an effective potential $u_{\rm{eff}}(r)$ and the \textcolor{black}{parameters (density, $u(r)$, and strength of self-propulsion relative to diffusion)} input into simulations of active particles. We note that a machine learning approach already exists to find activity parameters (one-body force) and $u(r)$ (two-body force), tested with ABPs and a variety of potentials.\cite{ruizgarcia2024} The choice of whether to find $u_{\rm{eff}}(r)$ or the true interaction $u(r)$ and activity parameters, depends on whether the primary interest is to understand the system’s structure or dynamics.

Effective pair potentials, $u_{\rm{eff}}(r)$, have been used to predict phase separation of active particles; these effective potentials have been obtained both by iterative Boltzmann inversion (IBI)\cite{trefz2016} and by deriving a relationship between $u(r)$ and $u_{\rm{eff}}(r)$ from the low-density limit.\cite{farage2015}  For weakly active systems, effective pair potentials have been obtained using an adaptation of IBI: the passive potential is used as input, and only the active contribution to $u_{\rm{eff}}(r)$ is found via IBI.\cite{evans2024} \textcolor{black}{We note that the definition of $u_{\rm{eff}}(r)$ used by \citet{evans2024} is the same as ours.} Iterative Ornstein-–Zernike Inversion, a modification of IBI, has been used to obtain effective interactions between granular particles –-- another inherently non-equilibrium system –-- again to describe phase separation and segregation, as in Ref.\cite{rodriguez2019}.

Inverse methods have been established for equilibrium suspensions of purely passive particles, to relate their structure –-- as described by the radial distribution function $g(r)$ --– to the underlying pair potential $u(r)$. Such methods, based on molecular simulation, include IBI\cite{reith2003} and inverse Monte Carlo.\cite{lyubartsev2017} They involve an initial guess for $u(r)$ followed by an update scheme. Other methods are based on machine learning.\cite{reeszimmerman2025,kampa2025} A particularly efficient form of inversion is based on test-particle insertion (TPI), with the method’s efficiency resulting from removing the need to rerun a Monte Carlo simulation at every iteration.\cite{reeszimmerman2025} In addition, not regenerating particle coordinates with further Monte Carlo simulation aids convergence to the correct potential.\cite{stones2019} As with IBI, TPI is a model-free inverse method, not requiring any assumptions about the form of $u(r)$. \textcolor{black}{The main limitations of TPI, also experienced with IBI, are difficulties at high density and requiring equilibrium.\cite{reeszimmerman2025}} \textcolor{black}{The test-particle insertion method involves finding the potential energy change associated with inserting hypothetical test-particles. With TPI, the difficulty at high density arises due to poor sampling of `successful' insertions.\cite{reeszimmerman2025b}}

This paper applies TPI to relate the structure of active particle suspensions to effective pair potentials. The inversion raises the question if there is an analogue of Henderson’s (1974)\cite{henderson1974} theorem: as long as the active particles have reached a steady-state structure, described by $g(r)$, should we expect a unique corresponding $u_{\rm{eff}}(r)$ at a given temperature and density? 

In cases where $u_{\rm{eff}}(r)$ can indeed be found, we subsequently measure thermodynamic properties –-- such as the pressure and chemical potential –-- of a system of passive particles interacting via $u_{\rm{eff}}(r)$, which we hope will shine light on questions around the construction of a statistical mechanical framework describing active matter.\cite{fodor2018} 

In what follows, we will outline the theory of the inverse method, the simulations of ABPs, and the potentials modelled in Sec.~\ref{sec:Theory}. The results of the inverse method as applied to ABPs are presented and discussed in Sec.~\ref{sec:results}, before drawing conclusions.

\section{Theory}\label{sec:Theory}
\subsection{Inverse method}\label{subsec:InverseMethod}
The radial distribution function, $g(r)$, is the ratio of the local number density about a reference particle, $\rho(r)$, and the bulk number density, $\rho$:
\begin{equation}\label{eq:1}
g_{\rm{\textcolor{black}{DH}}}(r)=\rho(r)/\rho,
\end{equation}
where $r$ is the position relative to the reference particle. Typically, $g(r)$ is found via the distance-histogram method \textcolor{black}{(denoted by the subscript DH)}, whereby the domain is divided into concentric rings of width $\Delta r$ about the reference particle, and the local number density in each ring evaluated.

An alternative method is based on test-particle insertion, where $g(r)$ is the ratio of the local to bulk ensemble average:
\begin{equation}\label{eq:2}
g\textcolor{black}{_{\rm{TPI}}}\left(r\right)=\frac{\langle\exp \lbrack-\beta\Psi \left(r\right) \rbrack \rangle}{\langle \exp \lbrack-\beta \Psi \rbrack \rangle}\textcolor{black}{.}
\end{equation}
\textcolor{black}{Here, we use the subscript TPI to denote that $g(r)$ is found by the test-particle insertion method. Also,} $\beta=1/k_{\rm{B}} T$, with $k_{\rm{B}} T$ being the thermal energy. \textcolor{black}{In the case of an equilibrium system,} $\Psi$ is the additional potential energy from the hypothetical insertion of a test particle.\cite{stones2019,widom1963} \textcolor{black}{Only considering pairwise interactions between particles,} 
\begin{equation}
\textcolor{black}{\Psi = \sum_{i=1}^{N} u(r_{ti}),} 
\end{equation}
\textcolor{black}{where $r_{ti}$ denotes the distance between the test particle $t$ and a real particle $i$, and $N$ is the number of particles in the system. In practice, we only consider particle--test-particle distances less than a cutoff $r_{\rm{c}}$ beyond which it is assumed that $\beta u(r)\sim0$.} In the numerator \textcolor{black}{of eq.~(\ref{eq:2})}, $\Psi(r)$ is measured with respect to a specific \textcolor{black}{particle--test-particle} distance $r$; \textcolor{black}{in the denominator,} $\Psi$ is no longer position dependent. \textcolor{black}{Note that, in our notation, $\rho$ and $\Psi$ refer to the reference density and potential, respectively, whereas $\rho(r)$ and $\Psi(r)$ refer to distance-dependent ones.}

In this work \textcolor{black}{on out-of-equilibrium ABP systems}, \textcolor{black}{instead of using the true thermodynamic potential $\Psi$, we adopt an \textit{effective} additional potential energy, $\Psi_{\rm{eff}}$. \textcolor{black}{In eq.~\ref{eq:2}, we replace $\Psi$ with $\Psi_{\rm{eff}}$.
We calculate our effective $\Psi_{\rm{eff}}= \sum_{i=1}^{N} u_{\rm{eff}}(r_{ti})$}} only \textcolor{black}{considering} effective pairwise interactions between particles. 

\textcolor{black}{For an equilibrium system, we note from eq.~(\ref{eq:2}) that $g_{\rm{TPI}}(r)$ is a function of $\beta\Psi(r)$, which is in turn a function of $\beta u(r)$. This leads to the formulation of an inverse method to find }the $\textcolor{black}{\beta} u(r)$ which makes the $g(r)$ obtained from each method match, \textcolor{black}{i.e., $g_{\rm{TPI}}(r)=g_{\rm{DH}}(r)$}. From Henderson’s uniqueness theorem, just one such $u(r)$ up to a constant should be obtained for a given structure, at a given temperature and density.\cite{henderson1974} \textcolor{black}{Extending this to steady-state ABP systems, we likewise find the $\beta u_{\rm{eff}}(r)$ that makes $g(r)$ from the two methods match.}

Here, we use the following approach: we begin an iterative predictor-corrector scheme with an initial guess,
\begin{equation}\label{eq:3}
\textcolor{black}{u_{\rm{eff,}0}} (r)=-k_{\rm{B}} T \ln \lbrack g_{\rm{DH}}(r)\rbrack,
\end{equation}
\textcolor{black}{where the index 0 denotes the 0\textsuperscript{th} iteration.} Then, the corrector proposed by Schommers (1973),\cite{schommers1973}
\begin{equation}\label{eq:4}
	\textcolor{black}{u_{\mathrm{eff,}{j+1}} (r)=u_{\mathrm{eff,}j} (r)}-k_{\rm{B}} T \ln{\left[ \frac{g_{\rm{DH}}(r)}{g_{\textcolor{black}{\mathrm{TPI}},j}(r)}\right]},
\end{equation}
is used to update the guess of \textcolor{black}{$u_{\mathrm{eff},j}(r)$} until convergence is achieved, where $g_{\textcolor{black}{\mathrm{TPI}},j}(r)$ is found by test-particle insertion and $j$ is the \textcolor{black}{iteration} number. 

\subsubsection{Convergence}\label{subsubsec:Convergence}
Convergence is assessed via the metric $\chi^2=\sum_{i}\left(g_{\rm{DH}} (r_i )-g_{\textcolor{black}{\mathrm{TPI}},j} (r_i)\right) ^2$, where $r$ has been discretised into bins with midpoints $r_i$. The initial guess [eq.~(\ref{eq:3})] corresponds to $u_{-1} (r)=0$ inputted into the right-hand side of eq.~(\ref{eq:4}); this is the potential of mean force, which we denote as $\textcolor{black}{w(r)}$.

Monte Carlo (MC) simulations (details in Appendix~A) of passive particles with the obtained $\beta u_{\rm{eff}}(r)$ are then run, and the obtained $g_{\rm{DH,MC}}(r)$ is compared with the $g_{\rm{DH}}(r)$ from the original simulations of the active particles. Convergence of the inverse method, and agreement between $g_{\rm{DH}}(r)$ and $g_{\rm{DH,MC}}(r)$, were checked for all examples in this paper, but are not shown for brevity.

The potential of mean force is expected to be a good description of dilute passive systems. For relatively dilute passive systems, this leads to fast convergence of the inverse method. Similarity between $u(r)$ and \textcolor{black}{$w(r)$} for an LJ passive system, at the same density, is shown in Appendix~C.  

For a dilute gas, the pair potential is equal to the potential of mean force, hence $g(r)=-\exp\lbrack\beta u(r) \rbrack$. For denser fluids, this can be considered the first term of an expansion of $g(r)$ in terms of number density.\cite{hunter2001} We write \textcolor{black}{$g(r) =-\exp\lbrack\beta w(r)\rbrack$, which is $g(r) =-\exp\lbrack\beta\left(u(r) + \textcolor{black}{u_{\rm{ind}}}(r)\right)\rbrack$,}  where we \textcolor{black}{have} split \textcolor{black}{$w(r)$} into:
\begin{equation}\label{eq:5}
 \textcolor{black}{w(r)}=u(r)+u_\mathrm{{\textcolor{black}{ind}}}(r), 	
 \end{equation}
with \textcolor{black}{$u_{\rm{ind}}(r)$} accounting for indirect interaction. We interpret some of the results for the effective pair potentials in this way, i.e., \textcolor{black}{$w_{\rm{eff}}(r)=u_{\rm{eff}}(r)+u_{\rm{ind,eff}}(r)$}.

\subsection{Simulations}\label{subsec:simulations}
\subsubsection{Potentials}\label{subsubsec:potentials}
The motion of ABPs combines passive interactions with an active component to their velocity. For the passive component, we model three different forms of interaction potential between particles: the Lennard-Jones (LJ), the Weeks–Chandler–Anderson (WCA) and the shoulder potential.

The LJ potential combines a steep short-range repulsion with an attractive tail\cite{lennardjones1931}; we truncate and shift the potential at $r=2.5\sigma$:
\begin{equation}\label{eq:6}
u_{\rm{LJ}}(r)=4\varepsilon\left[\left(\frac{\sigma}{r}\right)^{12}-\left(\frac{\sigma}{r}\right)^6 \right], 
\end{equation}
where $r$ is the centre-to-centre distance between particles, and $\varepsilon$ is the potential well depth, which we set as the energy scale. The WCA potential is a purely hard-core repulsive potential formed by truncating and shifting the Lennard-Jones potential to zero after a cutoff of $r=2^{1/6}\sigma$:\cite{weeks1971}
\begin{equation}\label{eq:7}
u_{\rm{WCA}}(r)=4\varepsilon\left[\left(\frac{\sigma}{r}\right)^{12}-\left(\frac{\sigma}{r}\right)^6 \right]+ \varepsilon.
\end{equation}
The shoulder potential has two length scales: a hard repulsive core (diameter $\sigma$) surrounded by a soft repulsive shell (diameter $\sigma_{\rm{s}}$, height $\varepsilon_{\rm{s}}$), truncated and shifted at $r=2.8\sigma$:
\begin{equation}\label{eq:8}
u_{\rm{s}}(r)=\varepsilon\left(\frac{\sigma}{r}\right)^{n}+\frac{1}{2} \varepsilon_{\rm{s}} \lbrack 1-\tanh \left(k_0 \left(r-\sigma_{\rm{s}} \right)\right)\rbrack.
\end{equation}
The stiffness of the core is described by $n$ and the steepness of the shoulder decay is described by $k_0$. This work takes $n=14$ and $k_0=10/\sigma$.\cite{gribova2009} All the examples in this work use $\sigma_{\rm{s}}=2.5\sigma$ and $\varepsilon_{\rm{s}}=2\varepsilon$.

\subsubsection{Active Brownian particles}\label{subsubsec:ABPs}
Trajectories of the coordinates of ABPs were obtained via 2D large-scale atomic/molecular massively parallel simulator (LAMMPS) simulation.\cite{plimpton1995} The periodic box size was adjusted to contain 2500 particles. The time evolution of the position, $\textcolor{black}{\bm{r}_i}$, and orientation relative to the \textit{x}-axis, $\theta_i$, of the $i$th active particle in an overdamped system is described by:
\begin{equation}\label{eq:9}
\textcolor{black}{\dot{\bm{r}}}_i=\frac{D_{\rm{t}}}{k_{\rm{B}} T} \left(-\sum_{j\neq i}\bm{\nabla} u\left(r_{ij} \right) + \textcolor{black}{F_{\rm{a}} \bm{n}_i} \right)+\sqrt{2D_{\rm{t}}} \textcolor{black}{\bm{\xi}_i},
\end{equation}
and
\begin{equation}\label{eq:10}
\dot{\theta}_i=\sqrt{2D_{\rm{r}}} \xi_{i,\theta}, 
\end{equation}
where $D_{\rm{t}}$ and $D_{\rm{r}}$ are the translational and rotational diffusion coefficients, respectively. We assume that $D_{\rm{t}}$ and $D_{\rm{r}}$ can be varied independently in an active system.\cite{martinroca2021} Activity arises via a self-propulsion force, $F_{\rm{a}}$, acting along the orientation vector $\textcolor{black}{\bm{n}_i}$, at angle ${\theta}_{i}$ to the \textit{x}-axis. The components of the thermal forces, ${\textcolor{black}{\bm{\xi}}_i}$ and $\xi_{i,\theta}$, are white noise with zero mean and correlations $\langle \xi_i^{\alpha}(t) \xi_j^{\beta}(t^{'})\rangle=\delta_{ij} \delta_{\alpha\beta} \delta\left(t-t^{'}\right)$, where $\alpha$ and $\beta$ are the \textit{x}- and \textit{y}- components, and $\langle\xi_{i,\theta} (t) \xi_{j,\theta} (t^{'})\rangle=\delta_{ij} \delta\left(t-t^{'}\right)$\textcolor{black}{, with $t$ denoting time}.

We vary the inputted $u(r)$ [eq.~(\ref{eq:9})], density and P\'eclet number, \textcolor{black}{$\mathrm{Pe}=3 F_{\mathrm{a}} M/\sigma D_{\rm{r}}$, where $M=D_{\mathrm{t}}/k_{\rm{B}} T$ is the mobility. The P\'eclet number} gives the ratio of advective active transport to diffusive transport\cite{martinroca2021}. \textcolor{black}{ In this work, $\mathrm{Pe}$ is changed by varying $F_{\mathrm{a}}$, with fixed $M$ and $D_{\mathrm{r}}$. We} obtain snapshots of the generated particle coordinates for each \textcolor{black}{combination}. It is ensured that the simulation has run for sufficient time to guarantee that the system is in steady state. We average over the steady-state snapshots and obtain $g_{\rm{DH}}(r)$. This is inverted according to Sec.~\ref{subsec:InverseMethod} to obtain $\beta u_{\rm{eff}}(r)$ for each set of input parameters.

The (reduced) number density is computed as $\rho=\sigma^2 N/A$, where $N\textcolor{black}{=2500}$ is the number of particles contained within the box area $A$. We simulated runs of length $10^7$ time steps, with a time step of $\Delta t/\tau =10^{-5}$, where $\tau=\sqrt{m\sigma^2/\epsilon}$, with $m$ being the mass of a particle. Every 10000\textsuperscript{th} frame is saved for analysis (corresponding to 1000 saved frames). Up to the first 200 of these saved frames are discarded, due to not yet being at steady state.  

\subsection{Calculations}\label{subsec:Calculations}
\subsubsection{Effective chemical potential}\label{subsubsec:EffChemPot}
\textcolor{black}{For a system to be in thermodynamic equilibrium, besides thermal and mechanical equilibrium, it is also required to be in chemical equilibrium. This is obtained when the chemical potential, $\mu$, of each species is constant throughout the system. For ABPs, it is known that there is no unique definition of state variables, such as pressure and temperature,\cite{solon2015,solon2015b} and similarly for chemical potential.\cite{paliwal2018}} \textcolor{black}{However, }test-particle insertion leads to a simple method to calculate \textcolor{black}{an} effective chemical potential, $\mu_{\rm{eff}}$:\cite{widom1963}
\begin{equation}\label{eq:11}
\mu_{\rm{eff}}=\mu^{\rm{o}} + k_{\rm{B}}T \ln {\rho} + \mu_{\rm{ex}},
\end{equation}
where we assume that there exists a reference chemical potential, $\mu^{\rm{o}}$. The excess contribution is given by
\begin{equation}\label{eq:12}
\mu_{\rm{ex}}=-k_{\rm{B}}T \ln\left(\langle\exp\lbrack-\beta\Psi_{\textcolor{black}{\rm{eff}}} \rbrack \rangle \right).	
\end{equation}
\textcolor{black}{This effective chemical potential is the chemical potential of a passive system with pairwise interactions described by $\beta u_{\rm{eff}}(r)$.} We note that the ensemble average $\langle\exp\lbrack-\beta\Psi_{\textcolor{black}{\rm{eff}}}\rbrack\rangle$ was already calculated for the denominator of eq.~(\ref{eq:2}) \textcolor{black}{(replacing $\Psi$ with $\Psi_{\rm{eff}}$ in the equation)}, and so we directly obtain $\beta \mu_{\rm{ex}}$ from $\beta u_{\rm{eff}}(r)$. The ensemble average is calculated as if we had an equilibrium system, with particles interacting via $u_{\rm{eff}}(r)$. Since the box size was kept the same throughout a given LAMMPS simulation, we follow the approach for finding a canonical ensemble average. 
We note that there exist alternative approaches to defining chemical potentials for active systems, constructed such that the gradients in the chemical potential correspond to flows.\textcolor{black}{\cite{hermann2019,paliwal2018,stenhammar2013,stenhammar2014}} In contrast, our definition of effective chemical potential is related to the system structure (not dynamics).

\subsubsection{Pressure}\label{subsubsec:pressure}
\paragraph{\textcolor{black}{Effective pressure.}}\label{para:EffPassivePressure}
The pressure of a passive system described by $u_{\rm{eff}}(r)$ can be calculated. This approach does not give the real pressure of the system (see Sec.~\ref{para:TotalPressure})\textcolor{black}{,} since it does not account for the active contribution to the pressure.\cite{sanoria2021}

The traditional method for calculating a pressure from a pair potential would be through the virial equation.\cite{allen1987} However, this would require calculating the derivative $\mathrm{d}u_{\rm{eff}}(r)/\mathrm{d}r$. Since the inversion only obtains binned values for $u_{\rm{eff}}(r)$, the calculated derivative will be noisy, leading to questions about the best smoothing and uncertainty in the calculated effective pressure, $P_{\rm{eff}}$.

We adopt an alternative method that does not involve calculating the virial, and so avoids the disadvantages associated with evaluating the derivative $\mathrm{d}u_{\rm{eff}}(r)/\mathrm{d}r$. The test-volume method requires evaluating the energy changes with hypothetical changes in the system volume, whereby the distances between particles are proportionally expanded or contracted.\cite{eppenga1984,harismiadis1996} In 2D, we seek the limit as $\Delta A\to0$ of
\begin{equation}\label{eq:13}
\beta P_{\rm{eff}}=\frac{N}{A}+\frac{1}{\Delta A}\ln{\left(\langle \exp{\left(-\beta\Delta U\right)}
   \rangle\right)},
\end{equation}
where $\Delta U$ is the difference in potential energy between the system at areas $A+\Delta A$ and $A$. The potential $U$ is computed as
\begin{equation}\label{eq:13b}
U=\frac{1}{2}\sum_{i}\sum_{j}u_{\rm{eff}}\left(r_{ij}\right).
\end{equation}
The ensemble average $\langle\exp{\left(-\beta\Delta U\right)}\rangle$ is found by averaging over the snapshots. Note that we use the particle coordinates as generated from the ABP simulations in this calculation --- as opposed to using a passive simulation with $u_{\rm{eff}}(r)$. \textcolor{black}{This is because the effective pressure is found by probing how the effective free energy changes with small changes in volume. This probes how $u_{\rm{eff}}(r)$ changes with small changes in $r$. However, we find a discretised $u_{\rm{eff}}(r)$, which has two consequences: (1) making $P_{\rm{eff}}$ sensitive to the discretisation (so we will present error bars for $P_{\rm{eff}}$ based on this), and (2) making the coordinates from a passive simulation with $u_{\rm{eff}}(r)$ not as accurate as the original coordinates. For the latter reason, we use the original ABP coordinates, to obtain as accurate a measure of $P_{\rm{eff}}$ as we can.}

\textcolor{black}{As further justification for finding the effective chemical potential as per our definition, the relationship between $P_{\rm{eff}}$ and $\mu_{\rm{eff}}$ is discussed in light of the Gibbs--Duhem equation in Appendix~D. Evaluating $\mu_{\rm{eff}}$ therefore offers another route to finding $P_{\rm{eff}}$, which might be less sensitive to discretisation. Moreover, we note that we find $\mu_{\rm{eff}}$ using the original ABP coordinates, since our inverse method naturally gives us $\Psi_{\rm{eff}}$ based on these coordinates. It is therefore consistent to also use these original ABP coordinates for the calculation of $P_{\rm{eff}}$.}

\paragraph{Total pressure.}\label{para:TotalPressure}
\textcolor{black}{The pressure of an active system can be used to probe its phase behaviour.\cite{sanoria2021,winkler2015}} Whilst there is some debate as to how to compute the pressure, $P$, for active systems,\cite{solon2015} we adopt
\begin{equation}\label{eq:14}
P=\frac{Nk_{\rm{B}} T}{A}+\frac{\textcolor{black}{ F_{\rm{a}}}}{2A D_{\rm{r}}}  \sum_{i=1}^{N} \langle \textcolor{black}{\bm{n}_i}\cdot \textcolor{black}{\bm{r}_i}\rangle +\frac{1}{4A} \sum_{i=1}^{N}\sum_{j=1}^{N}\langle \textcolor{black}{\bm{F}_{ij}}\cdot \textcolor{black}{\bm{r}_{ij}} \rangle ,	\end{equation}
where here the force between particles $i$ and $j$, $\textcolor{black}{\bm{F}_{ij}}$, acts along $\textcolor{black}{\bm{r}_{ij}}$, such that $\textcolor{black}{\bm{F}_{ij}} \cdot \textcolor{black}{\bm{r}_{ij}}=-  \mathrm{d}u(r)/\mathrm{d}r\rvert_{r_{ij}} r_{ij}$.\cite{martinroca2022} \textcolor{black}{The first term is the ideal contribution.} The second term is the contribution from activity, termed the `swimming pressure'.\textcolor{black}{\cite{takatori2014,omar2020}} The final term is the contribution from the interaction potential (i.e., the usual virial term). \textcolor{black}{Our main aim is to acquire a measurement of pressure for the ABPs, in order to compare with our effective pressures. We adopt the definition that we have used in previous work.\cite{martinroca2022}}

\section{Results and discussion}\label{sec:results}
\subsection{Effective pair potentials}\label{subsec:EffPairPot}

A series of systematic studies are carried out, varying (1) the underlying passive pair potential (for the shoulder, WCA and LJ potentials), (2) P\'eclet number (for the WCA potential), and (3) density (for the shoulder potential). The particle number density was set to $\rho=0.3$ in all cases, except for the study varying density. This was chosen to be sufficiently dense to show some structure, and not too dense as to pose difficulty for test-particle insertion. \textcolor{black}{Example snapshots from both the LAMMPS (i.e, molecular dynamics, MD) simulations and the MC simulations run using the obtained $\beta u_{\rm{eff}}(r)$ are presented for interest in Fig.~\ref{fig:snapshots} (a subset) and in Appendix~G (full set). Visually good agreement is obtained in the structure of the MD and MC snapshots, as expected, since $g_{\rm{DH}}(r)$ has been regenerated with low error. Differences in higher-order structural correlations are unlikely to be visible by eye.}

\begin{figure*}[t!]
    \centering
    \includegraphics[width=\linewidth]{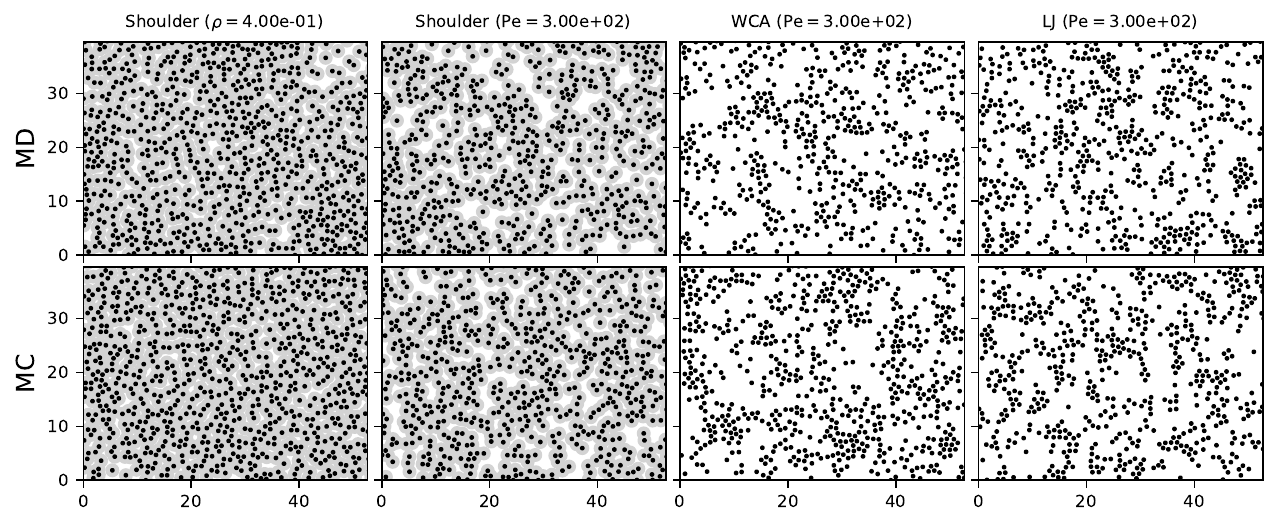}
    \caption{\textcolor{black}{Zoomed snapshots of the MD (top row) and MC (bottom row) simulations for the studied potentials for selected densities and P\'eclet numbers. From left to right: From the shoulder potential study at varying $\rho$, we present $\rho=0.4$ (at $\rm{Pe}=120$, see Fig.~\ref{fig:4}). Then, for the study varying potential type, we present the shoulder, WCA and LJ potentials (all at $\rho=0.3$ and $\rm{Pe}=300$, see Fig.~\ref{fig:1}). (Snapshots for the complete set of studied parameters are shown in Appendix~G.)}}
    \label{fig:snapshots}
\end{figure*}

\subsubsection{Varying potential types}\label{subsubsec:VaryingPot}
\begin{figure}[h!]
\centering
   \begin{subfigure}[b]{0.4\textwidth}
     \subcaption{}
    \includegraphics[width=1\textwidth]{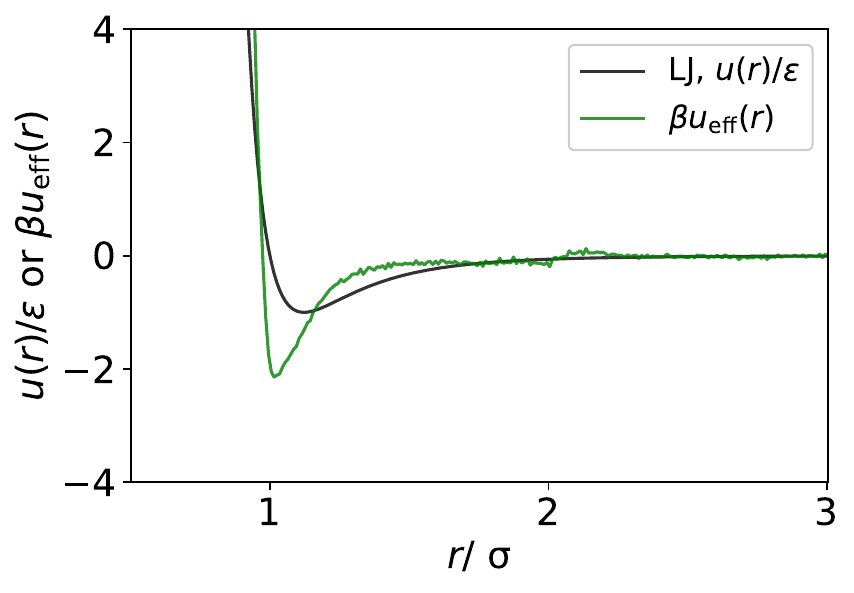}
        \label{fig:1a}
    \end{subfigure}
      \quad
         \begin{subfigure}[b]{0.4\textwidth}
     \subcaption{}
    \includegraphics[width=1\textwidth]{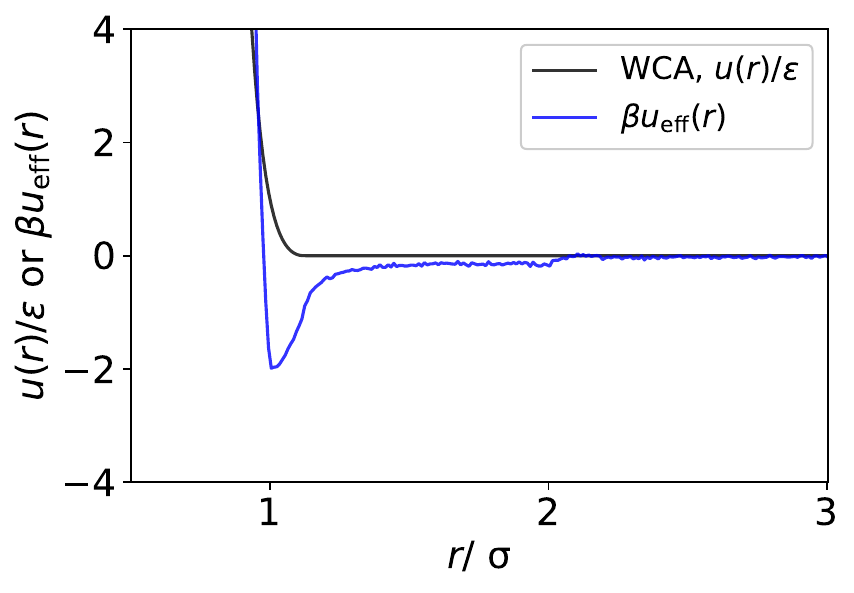}
        \label{fig:1b}
    \end{subfigure}
      \quad
         \begin{subfigure}[b]{0.4\textwidth}
     \subcaption{}
    \includegraphics[width=1\textwidth]{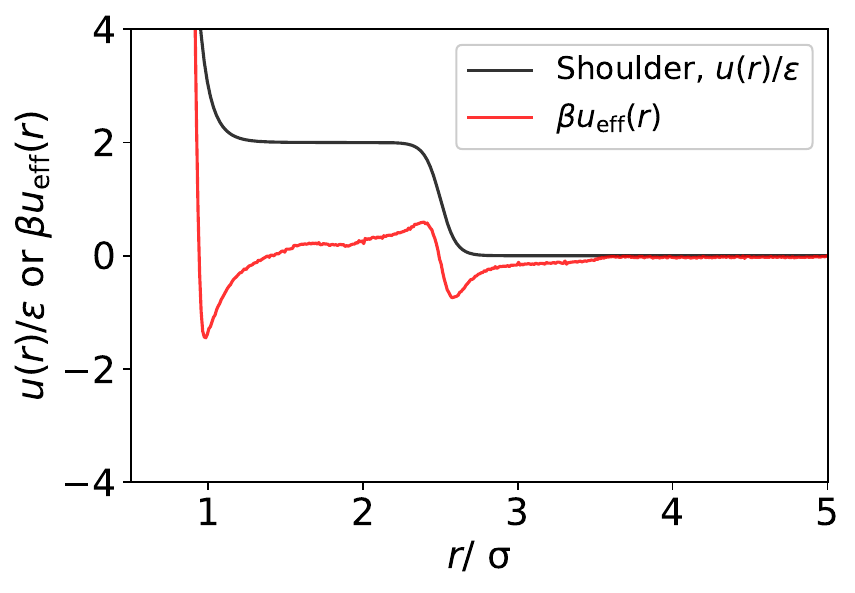}
        \label{fig:1c}
    \end{subfigure}
  \caption{Varying potential type, with fixed $\rm{Pe}=300$ \& $\rho=0.3$, for (\subref{fig:1a}) LJ, (\subref{fig:1b}) WCA and (\subref{fig:1c}) shoulder potentials. Each plot shows the passive potential $u(r)$ and $\beta u_{\rm{eff}}(r)$ from the inverse method.}
  \label{fig:1}
\end{figure}

We first observe that the inverse method successfully finds $\beta u_{\rm{eff}}(r)$ with different potential types: LJ, WCA and shoulder potentials. For each potential, Fig.~\ref{fig:1} compares the passive $u(r)$ used in the LAMMPS simulations [eq.~(\ref{eq:9})] with $\beta u_{\rm{eff}}(r)$ obtained by the inverse method. 

Qualitatively similar shapes are obtained for $u_{\rm{eff}}(r)$ for the WCA potential as for the LJ potential. \textcolor{black}{In part, this is because at $\rm{Pe}=300$, the activity is much more important than the underlying passive interactions.\cite{chacon2022}} \textcolor{black}{However, this similarity} is interesting, since WCA $u(r)$ (Fig.~\ref{fig:1b}) is entirely repulsive, whilst LJ $u(r)$ (Fig.~\ref{fig:1a}), in addition to its short-range repulsion, includes an attractive tail. We conclude that the activity is responsible for the attractive well in $u_{\rm{eff}}(r)$ for the WCA potential, resulting in similarity with the LJ potential’s $u_{\rm{eff}}(r)$. This attractive well in $u_{\rm{eff}}(r)$ was also found by Ref.\cite{evans2024}, which used a pseudo-hard sphere $u(r)$. \citet{evans2024} find that weak activity promotes percolation, even before MIPS takes place. 
For both LJ and WCA, we observe small jumps in $u_{\rm{eff}}(r)$ around $r=2\sigma$. From corresponding plots of $g(r)$ (see Appendix~B), which have a secondary peak around $r=2\sigma$, it appears that these bumps are merely from the structure.\cite{jeanneret2014} Activity must be responsible for the slight long-range repulsion in $u_{\rm{eff}}(r)$ around $r=2.2\sigma$ for the LJ system.\cite{evans2024} The presence of this jump in the shoulder potential’s $u_{\rm{eff}}(r)$ might be hidden by the two length scales in $u(r)$.

Interestingly, whilst the shoulder potential is entirely repulsive, $u_{\rm{eff}}(r)$ shows two attractive wells, around $r=\sigma$ and $r=\sigma_{\rm{s}}$. These effective attractive regions must be due to the activity --- an emergent phenomenon of repulsive ABPs.\cite{martinroca2021} \textcolor{black}{For passive systems, multi-well potentials have been found to give rich assembly results in simulations, in both 2D and 3D.\cite{engel2007,dshemuchadse2021} We have previously mapped a phase diagram for 2D shoulder potential ABPs.\cite{martinroca2022} The phase diagram has MIPS and homogeneous regions. Interestingly, within each phase, a rich range of structures is observed, including monomers, dimers, trimers, chains and hexagonal structures. This richness arises from the shoulder potential having two length scales --- which manifests as two attractive wells in $u_{\rm{eff}}(r)$.
}

Appendix~B demonstrates the convergence of the inverse method for each potential type, additionally showing the very good agreement between $g_{\rm{DH}}(r)$ from the original LAMMPS simulation and $g_{\rm{DH,MC}}(r)$ from MC simulations of passive particles with the obtained $\beta u_{\rm{eff}}(r)$. This is a key result: it shows that $\beta u_{\rm{eff}}(r)$ is able to regenerate the structure, and so, in this respect, is truly an effective passive potential for the active Brownian particles. It is an interesting result in itself that there is a $\beta u_{\rm{eff}}(r)$ that regenerates $g(r)$ in an equilibrium Monte Carlo simulation. Furthermore, we consider what thermodynamic properties, such as the pressure, might be encompassed in $\beta u_{\rm{eff}}(r)$.\cite{sanoria2021} This is explored in \textcolor{black}{Secs.~\ref{subsec:EffChemPot} and~\ref{subsec:Pressure}}.

\subsubsection{Varying the P\'eclet number}\label{subsubsec:VaryingPe}
The effect of varying P\'eclet number is also studied: this shows the influence of varying activity. 
\begin{figure}[h!]
 \centering
 \hspace{0.1 cm} \begin{subfigure}[b]{0.4\textwidth}
     \subcaption{}
    \hspace{0.04cm}\includegraphics[width=1\textwidth]{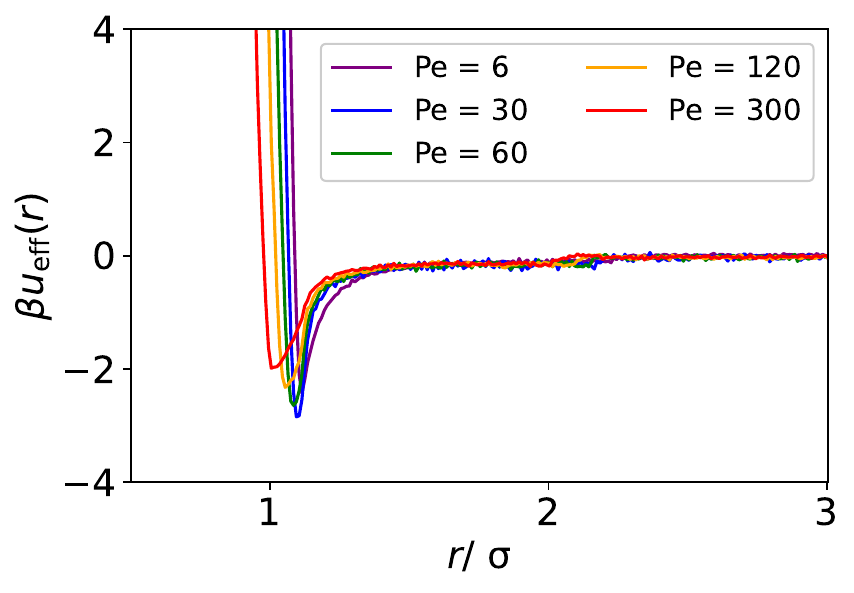}
        \label{fig:2a}
    \end{subfigure}
      \quad
       \begin{subfigure}[b]{0.4\textwidth}
     \subcaption{}
    \includegraphics[width=1\textwidth]{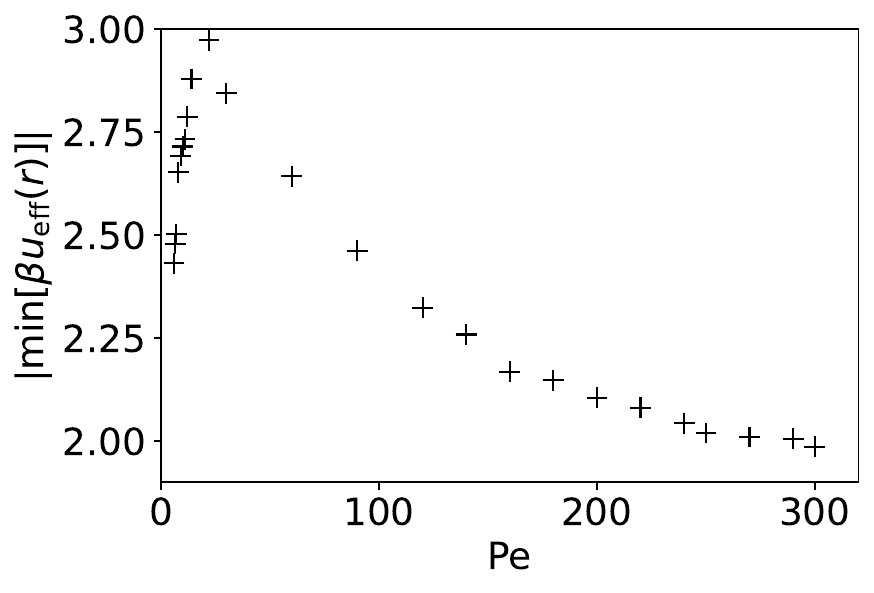}
        \label{fig:2b}
    \end{subfigure}
 \caption{Varying P\'eclet number for the WCA potential, with fixed $\rho=0.3$: (\subref{fig:2a}) Plot of $\beta u_{\rm{eff}}(r)$ at different $\rm{Pe}$. (\subref{fig:2b}) Plot of the magnitude of $\beta u_{\rm{eff}}(r)$’s well depth as a function of $\rm{Pe}$.}
 \label{fig:2}
\end{figure}

For the \textit{shoulder potential}, at P\'eclet numbers lower than 120, particles do not have enough energy to penetrate the soft repulsive shell, and so they tend to behave as particles with larger diameter ($\sigma_{\rm{s}}=2.5\sigma$). The effective packing fraction based on this larger diameter is greater than one, leading to ‘frustrated’ structures.\cite{martinroca2022} This results in particles forming snake-like paths of high local density. For the \textit{LJ} potential, at P\'eclet numbers lower than 300, there was too much particle clustering --- i.e., again, areas of local high density. This was reflected in peaks in $g(r)$ of high magnitude (e.g., $>10$).  These areas of high local density mean that test-particle insertion does not work well. This is due to the difficulty of ‘successfully’ inserting a particle in a region that is already very dense.\cite{boulougouris1999} This leads to either convergence not being achieved, or, if  achieved, a noisy $\beta u_{\rm{eff}}(r)$.  

We therefore choose to present the effect of the P\'eclet number using the WCA potential (see \textcolor{black}{Fig.~\ref{fig:2}}), which does not have the above problems and allows the study of a range of $\rm{Pe}$. Note that there is no universal density limit above which test-particle insertion fails; it depends on the effective potential. \textcolor{black}{In Appendix~E, we discuss the second virial coefficient and what that might tell us about the phase behaviour of the ABPs.}

Plots of $\beta u_{\rm{eff}}(r)$ for the WCA potential with a selection of $\rm{Pe}$ numbers ($\rm{Pe}=6,\ 30,\  60,\ 120\ \&\ 300$) at $\rho=0.3$ are shown in Fig.~\ref{fig:2a}. For the WCA potential at each $\rm{Pe}$ tested, excellent agreement between $g_{\rm{DH}}(r)$ and $g_{\rm{TPI}}(r)$ is obtained (not shown). We learn that the inverse method is able to work for different $\rm{Pe}$. We additionally plot the magnitude of the well depths in $u_{\rm{eff}}(r)$ as a function of further $\rm{Pe}$ values in Fig.~\ref{fig:2b}. 

The magnitude of the well depths in $u_{\rm{eff}}(r)$ are observed to initially increase as $\rm{Pe}$ is increased (from $\rm{Pe}=6$ to $\rm{Pe}=22$). However, at higher $\rm{Pe}$ ($\rm{Pe}>22$), the magnitude of the well depths in $u_{\rm{eff}}(r)$ decreases as $\rm{Pe}$ is increased. Whilst the apparent attraction in $u_{\rm{eff}}(r)$ is attributed to the activity, increasing activity appears to decrease the particle clustering. This is because ABPs tend to slow at high density for steric reasons, but increased activity reduces the effect of this slowing down.\cite{su2023, evans2024} In the limit of $\rm{Pe}\to0$, we would recover the $u(r)$ for a purely passive simulation of the shoulder potential. \textcolor{black}{In Appendix~F, we present how the mean free time between particle collisions varies with $\rm{Pe}$, finding similar trends to those observed with $u_{\rm{eff}}(r)$.}

\begin{figure}[h!]
 \centering
 \begin{subfigure}[b]{0.4\textwidth}
 \subcaption{}
   \includegraphics[width=1\textwidth]{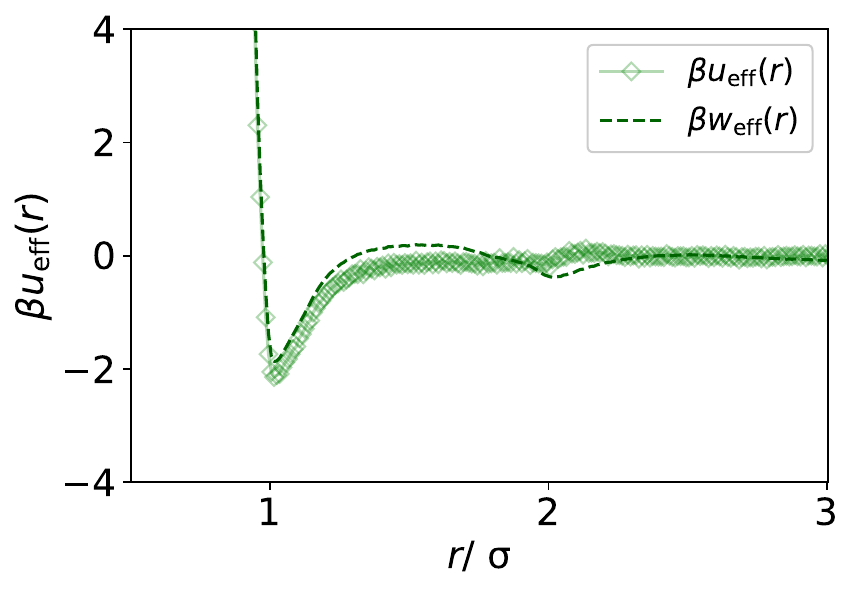}
        \label{fig:3a}
    \end{subfigure}
          \quad
     \begin{subfigure}[b]{0.4\textwidth}
      \subcaption{}
    \hspace{0.22 cm}\includegraphics[width=1\textwidth]{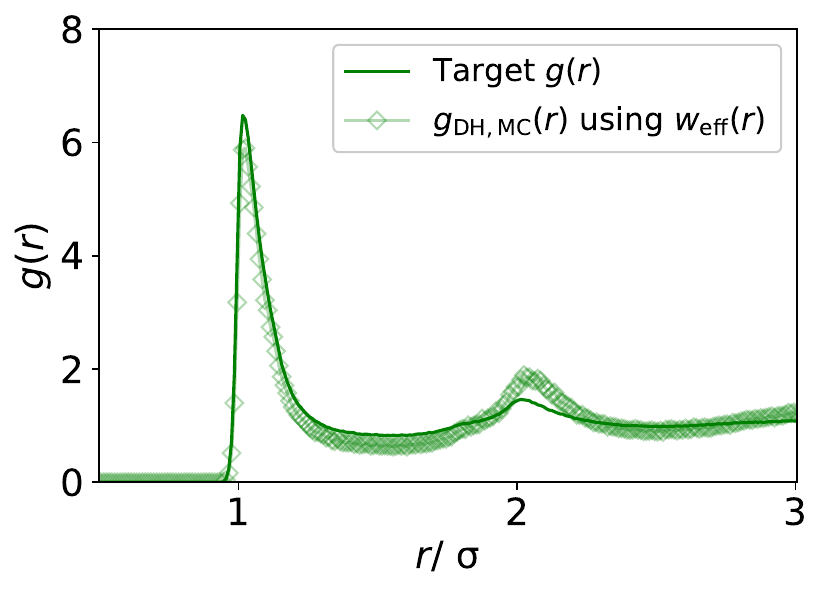}
        \label{fig:3b}
    \end{subfigure}
      \quad
 \caption{Demonstration of the closeness of $\beta \textcolor{black}{w_{\rm{eff}}(r)}$ to $\beta u_{\rm{eff}}(r)$, using the example of ABPs with a LJ potential, with $\rm{Pe}=300$ \& $\rho=0.3$: (\subref{fig:3a}) Comparison of $\beta \textcolor{black}{w_{\rm{eff}}(r)}$ with $\beta u_{\rm{eff}}(r)$. (\subref{fig:3b}) Comparison of $g_{\rm{DH}}(r)$ from the original data with $g_{\rm{DH,MC}}(r)$ from a MC simulation using $\beta \textcolor{black}{w_{\rm{eff}}(r)}$. \textcolor{black}{These $g(r)$ should not agree perfectly, as they correspond to different $u(r)$, though they are similar due to the closeness of $\beta \textcolor{black}{w_{\rm{eff}}(r)}$ to $\beta u_{\rm{eff}}(r)$.}}
 \label{fig:3}
\end{figure}

\subsubsection{Effective potential of mean force}\label{subsubsec:EffPMF}
Taking the case of ABPs with a LJ potential shown in Fig.~\ref{fig:1a}, the effective potential of mean force, \textcolor{black}{$w_{\rm{eff}}(r)$} [eq.~(\ref{eq:3})], is plotted in Fig.~\ref{fig:3a}, as an example to compare with $u_{\rm{eff}}(r)$. The potentials \textcolor{black}{$w_{\rm{eff}}(r)$} and $u_{\rm{eff}}(r)$ appear very similar; this contributes to the rapid convergence of the inverse scheme when using \textcolor{black}{$w_{\rm{eff}}(r)$} as an initial guess.\cite{moore2014} In Fig.~\ref{fig:3b}, we compare the target $g_{\rm{DH}}(r)$ found from the original data, with that from an MC simulation using \textcolor{black}{$w_{\rm{eff}}(r)$}. The similarity between these $g(r)$ reflects that between $u_{\rm{eff}}(r)$ and \textcolor{black}{$w_{\rm{eff}}(r)$}. \textcolor{black}{The first peak in the target $g_{\rm{DH}}(r)$ is well reproduced. The second peak is higher in $g_{\rm{DH,MC}}(r)$ (generated using \textcolor{black}{$w_{\rm{eff}}(r)$}) than in $g_{\rm{DH}}(r)$, as a result of the second well in \textcolor{black}{$w_{\rm{eff}}(r)$}. A third peak appears in $g_{\rm{DH,MC}}(r)$ which is absent in $g_{\rm{DH}}(r)$, as higher-order structural correlations are more sensitive to small differences in the interaction potential.}

Therefore, it seems to be the case for relatively dilute active Brownian particles, that the effective potential of mean force explains most of the structure. The example in Fig.~\ref{fig:3} is at high $\rm{Pe}$ ($\rm{Pe}=300$); i.e., activity dominates the particles’ motion. We learn that, for active Brownian particles at $\rho=0.3$, effective direct interactions, which are predominantly a result of the activity, contribute most to the effective pair potential. Interestingly, activity indirectly leads to structure, but emerges as a direct contribution to the effective pair potential.

\subsubsection{Varying the density}\label{subsubsec:VaryingDensity}
Finally, the effect of varying particle density is examined, using the shoulder potential as an example, at $\rm{Pe}=120$. 

\begin{figure}[h!]
 \centering
 \hspace{0.18 cm}\begin{subfigure}[b]{0.4\textwidth}
 \subcaption{}
    \includegraphics[width=1\textwidth]{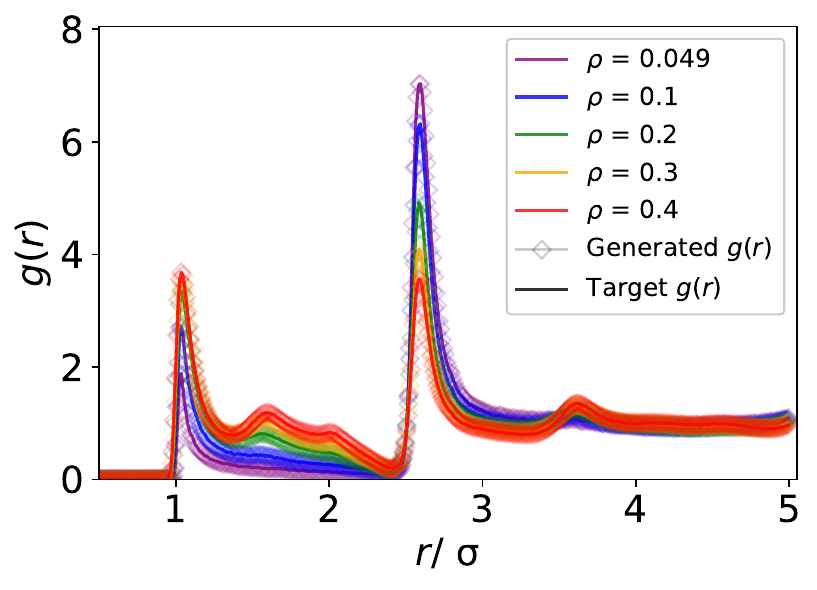}
        \label{fig:4a}
    \end{subfigure}
          \quad
     \begin{subfigure}[b]{0.4\textwidth}
 \subcaption{}
    \includegraphics[width=1\textwidth]{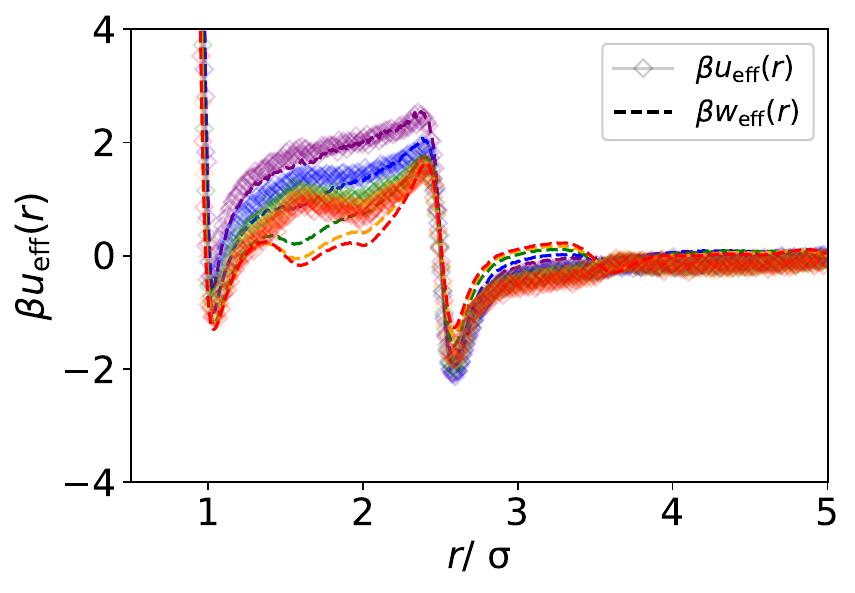}
        \label{fig:4b}
    \end{subfigure}
 \caption{Shoulder potential at varying density, with fixed $\rm{Pe}=120$. (\subref{fig:4a}) Plot comparing $g_{\rm{DH}}(r)$ and $g_{\rm{TPI}}(r)$, based on the converged $\beta u_{\rm{eff}}(r)$ from the inverse method. (\subref{fig:4b}) Plot comparing $\beta u_{\rm{eff}}(r)$ with $\beta \textcolor{black}{w_{\rm{eff}}(r)}$. The colour code in the key of (\subref{fig:4a}) also applies to (\subref{fig:4b}).}
 \label{fig:4}
\end{figure}

The shoulder potential is an interesting example because of its two length scales. The results are presented in Fig.~\ref{fig:4}. This is a particularly relevant study, as neither the $u(r)$ or activity change in the LAMMPS simulations of the ABPs. Nevertheless, we observe differences in $u_{\rm{eff}}(r)$ purely due to change in density. As the density is increased, the ratio of the peak in $g(r)$ around $r=\sigma$ to that around $r=\sigma_{\rm{s}}$ increases. This suggests that, at high density, more particles are forced within the soft-shell region due to steric restrictions. The density-dependence of $g(r)$ is unsurprising, since this is also observed in passive systems, due to structure increasing as density increases.\cite{thorneywork2014} However, in passive systems, the $u(r)$ obtained from the inversion method clearly does not depend on the density --– since this $u(r)$ should have converged to the original $u(r)$ used in the Monte Carlo simulation that generated the particle coordinates. In contrast, the $u_{\rm{eff}}(r)$ obtained for the ABPs do depend on the density, since the emergent effects of activity depend on the density. This is via density-dependent particle velocities: whilst density-dependence of the particle velocities is not directly coded into the model, collisions are expected to slow particles down at high density.\cite{cates2015,moran2022} We see that the activity leads to non-pairwise additivity.

Looking at $u_{\rm{eff}}(r)$, at all densities, attractive wells appear at the two characteristic distances. The depth of the well around $r=\sigma$ increases slightly with increasing density. More generally, considering the region between the two wells, the apparent attractiveness of the interactions increases with increasing density. This corresponds with the increased penetration of the soft-shell observed in $g(r)$ with increasing density. The depth of the second attractive well, around $r=\sigma_{\rm{s}}$, is less affected by the density.

Fig.~\ref{fig:4b} additionally shows $\textcolor{black}{w_{\rm{eff}}(r)}$ at each density. This makes it clear that similarity between $u_{\rm{eff}}(r)$ and $\textcolor{black}{w_{\rm{eff}}(r)}$, as for passive systems, decreases with increasing density. At $\rho=0.049$, there is very good agreement between $u_{\rm{eff}}(r)$ and $\textcolor{black}{w_{\rm{eff}}(r)}$. As the density increases, the agreement worsens because $w_{\rm{eff}}(r)$ changes more with increasing $\rho$ than $\textcolor{black}{u_{\rm{eff}}(r)}$ does. This reflects our understanding that effective indirect interactions \textcolor{black}{$u_{\rm{ind,eff}}(r)$} increase with increasing density.

\subsection{Effective chemical potential}\label{subsec:EffChemPot}
\begin{figure}[h!]
 \centering
  \begin{subfigure}[b]{0.45\textwidth}
 \subcaption{}
 \includegraphics[width=1\textwidth]{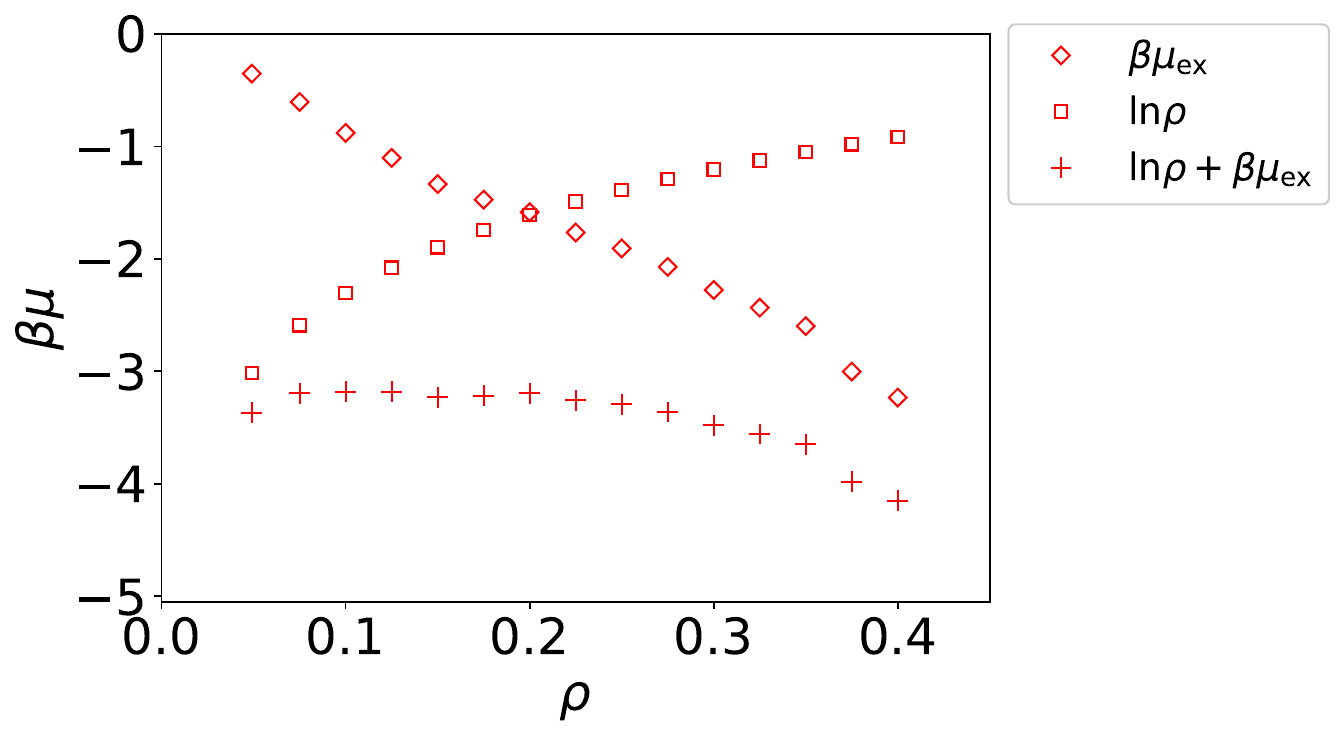}
        \label{fig:5a}
    \end{subfigure}
    \quad
    \begin{subfigure}[b]{0.45\textwidth}
 \subcaption{}
    \includegraphics[width=1\textwidth]{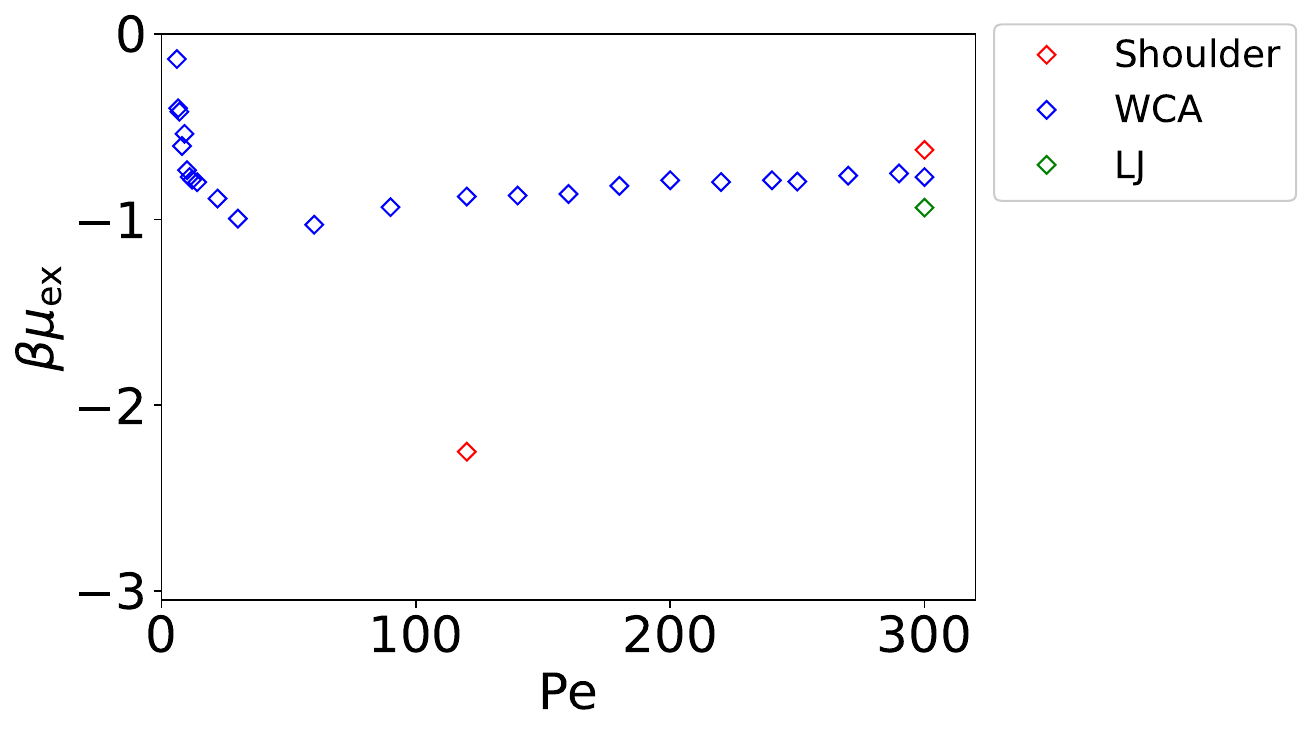}
        \label{fig:5b}
    \end{subfigure}
 \caption{Plots of effective chemical potential as a function of (\subref{fig:5a}) $\rho$ (shoulder potential with $\rm{Pe}=120$) and (\subref{fig:5b}) Pe (shoulder, WCA and LJ potentials, $\rho=0.3$).}
 \label{fig:5}
\end{figure}

Following the calculation method given in Sec.~\ref{subsubsec:EffChemPot}, we plot either $\beta(\mu_{\rm{eff}}-\mu^{\rm{o}} )=\ln{\rho}+\beta\mu_{\rm{ex}}$ (Fig.~\ref{fig:5a}, varying $\rho$) or $\beta\mu_{\rm{ex}}$ (Fig.~\ref{fig:5b}, fixed $\rho$) for the ABPs.

Considering the shoulder potential at varying density, Fig.~\ref{fig:5a} shows that the ideal chemical potential increases with density, whilst the effective excess chemical potential decreases; the overall effective chemical potential shows a maximum.

At fixed density, $\ln{\rho}$ is constant, so Fig.~\ref{fig:5b} plots the effective excess chemical potential only. For the WCA potential, there is a non-monotonic trend in the effective chemical potential as $\rm{Pe}$ is varied; this likely reflects the non-monotonic behaviour in the well depths in $u_{\rm{eff}}(r)$ (Fig.~\ref{fig:2b}). In addition to the WCA potential, Fig.~\ref{fig:5b} also plots results for the LJ and shoulder potentials. The effective chemical potentials for the LJ, WCA and shoulder potentials at $\rm{Pe}=300$ are close to each other. This is expected: the passive pair potentials become less important as the activity is increased, and the systems with the different potentials become increasingly similar to each other. For the shoulder potential, the effective chemical potential is higher at $\rm{Pe}=300$ than at $\rm{Pe}=120$; higher chemical potential corresponds with a less negative test-particle insertion energy [when averaged as in eq.~(\ref{eq:12})].

\subsection{Pressure}\label{subsec:Pressure}
\subsubsection{\textcolor{black}{Effective pressure}}\label{subsubsec:EffPassivePressure}
\begin{figure*}[h!]
 \centering
 \begin{subfigure}[b]{0.35156\textwidth}
 \subcaption{}
    \includegraphics[width=1\textwidth]{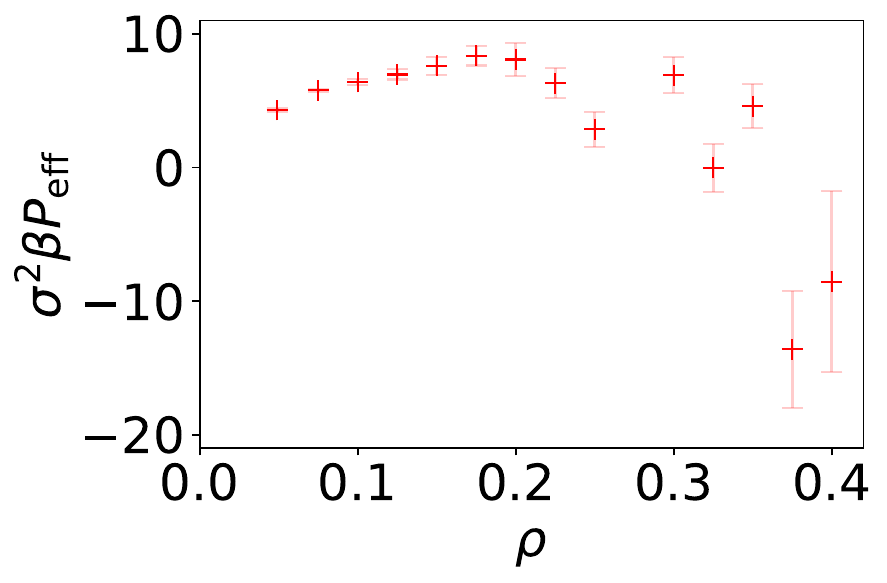}
        \label{fig:6a}
    \end{subfigure}
    \quad
     \begin{subfigure}[b]{0.54845\textwidth}
 \subcaption{}
    \includegraphics[width=1\textwidth]{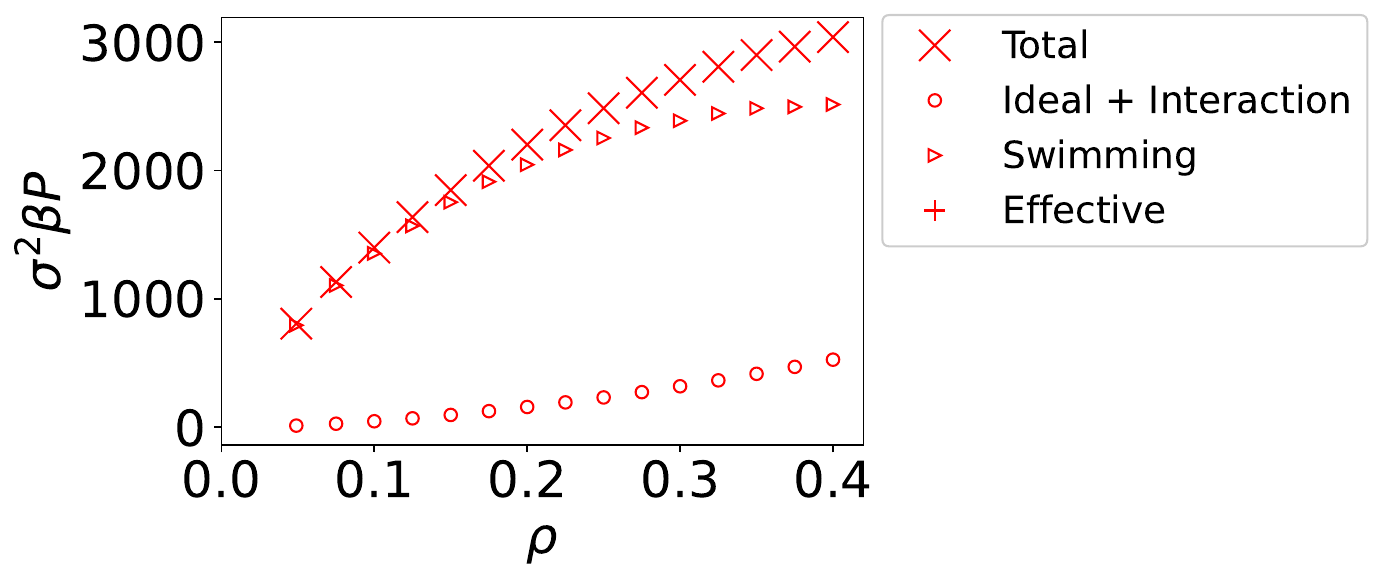}
        \label{fig:6b}
    \end{subfigure}
    \quad
     \begin{subfigure}[b]{0.34429\textwidth}
 \subcaption{}
 \hspace{0.14cm}  \includegraphics[width=1\textwidth]{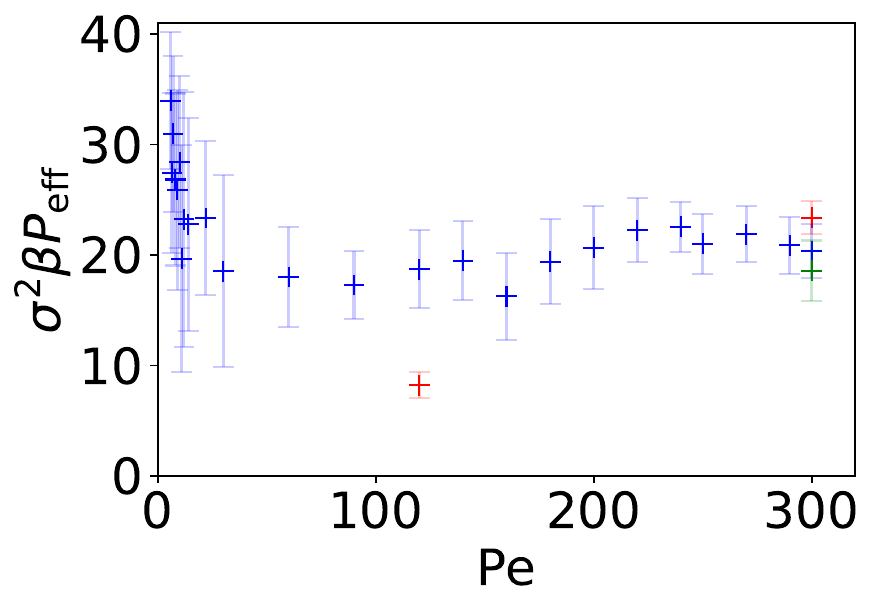}
        \label{fig:6c}
    \end{subfigure}
    \quad
     \begin{subfigure}[b]{0.55571\textwidth}
 \subcaption{}
     \includegraphics[width=1\textwidth]{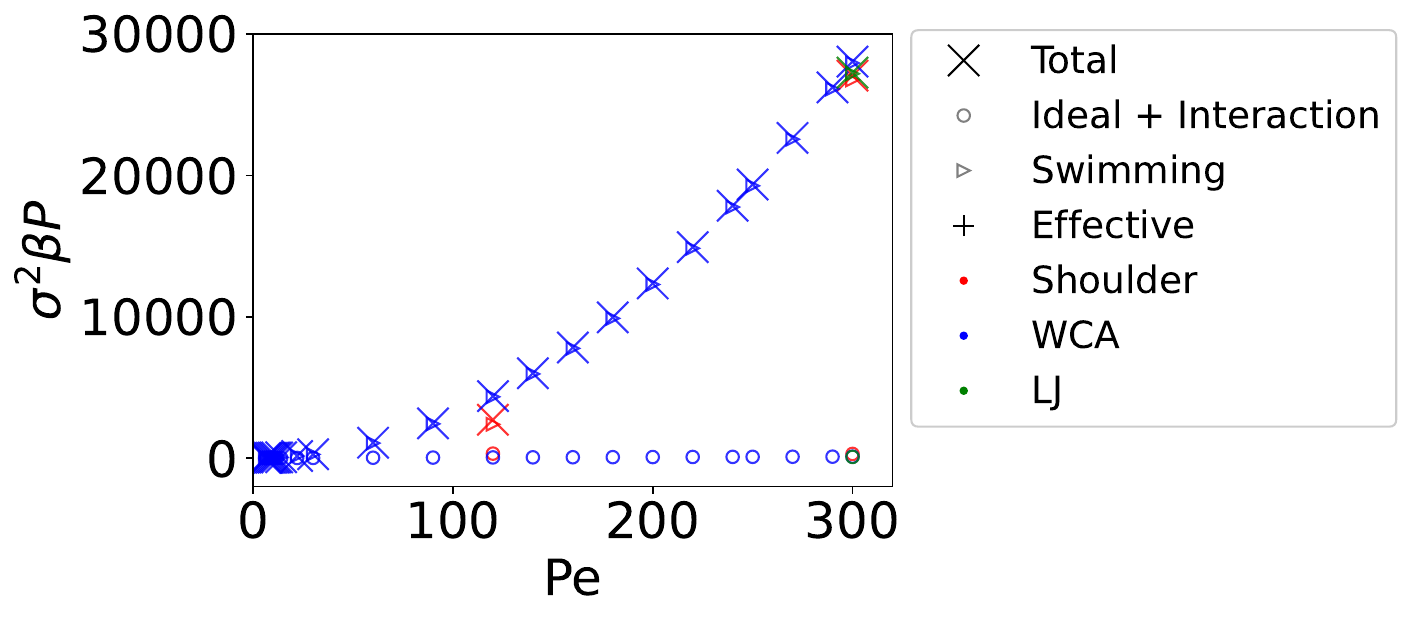}
        \label{fig:6d}
    \end{subfigure}
 \caption{Effective pressure with error bars [(\subref{fig:6a}) \& (\subref{fig:6c})], and total pressure\textcolor{black}{, divided into its components of `Ideal + Interaction' and `Swimming' pressures} [(\subref{fig:6b}) \& (\subref{fig:6d})]; both as a function of $\rho$ [shoulder potential with $\rm{Pe}=120$, (\subref{fig:6a})–(\subref{fig:6b})] and $\rm{Pe}$ [shoulder, WCA and LJ potentials, $\rho=0.3$, (\subref{fig:6c})–(\subref{fig:6d})]. The key in (\subref{fig:6b}) also applies to subfigure (\subref{fig:6a}), and, likewise, the key in (\subref{fig:6d}) also applies to subfigure (\subref{fig:6c}).}
 \label{fgr:example2col}
\end{figure*}

In this section, using eq.~(\ref{eq:13}), we find the pressures of passive systems with the same $u_{\rm{eff}}(r)$ as the ABPs \textcolor{black}{(using the ABP coordinates, as explained in Sec.~\ref{subsubsec:pressure})}. The results are presented in Figs.~\ref{fig:6a} \& \ref{fig:6c}. Strictly speaking, $u_{\rm{eff}}(r)$ is a function of $\rho$. Whilst this is ignored in calculating $P_{\rm{eff}}$, the result should be reasonably accurate, since the density-dependence of $u_{\rm{eff}}(r)$ is not too large (see Fig.~\ref{fig:4b}).\cite{louis2002} As with the effective chemical potential, the \textcolor{black}{effective pressure} appears similar between the LJ, WCA and shoulder potentials at high $\rm{Pe}$ ($\rm{Pe}=300$). For the shoulder potential at varying densities (Fig.~\ref{fig:6a}), the \textcolor{black}{effective pressure} shows a maximum, as did the effective chemical potential (Fig.~\ref{fig:5a}). Error bars are drawn in Figs.~\ref{fig:6a} \& \ref{fig:6c} based on the standard deviation of results from different bin widths. Due to the large error bar, the sign of $P_{\rm{eff}}$ at $\rho=0.4$ is uncertain. We stress that there is still less variance with bin width using this method than with the virial method.

For a passive system, the Gibbs--Duhem equation at constant temperature can be written as $\mathrm{d}\mu/\mathrm{d}\rho=(1/\rho)\mathrm{d}P/\mathrm{d}\rho$.\cite{barrat2003, hermann2019}   
By comparing the data shown in Figs.~\ref{fig:5a} \& \ref{fig:5b} with those in \ref{fig:6a} \& \ref{fig:6c}, Appendix~D attempts to test the Gibbs--Duhem equation, both for varying $\rho$ (shoulder potential) and $\rm{Pe}$ (WCA potential). However, we observe in Figs.~\ref{fig:2} \& \ref{fig:4} that $u_{\rm{eff}}(r)$ varies with $\rho$ and $\rm{Pe}$. \citet{louis2002} explains the challenges of deriving different thermodynamic properties, such as $\mu_{\rm{eff}}$ and $P_{\rm{eff}}$, from density-dependent effective pair potentials. Whilst we expect eq.~(\ref{eq:12}) for $\mu_{\rm{eff}}$ to still be exact, due to its similarity with the method used to find $u_{\rm{eff}}(r)$ in the first place, we noted above that this is not the case with our calculation of $P_{\rm{eff}}$. In future work, the role of many-body effects should be considered, in determining whether the Gibbs--Duhem relationship holds for the effective thermodynamics of ABPs. Nevertheless, $u_{\rm{eff}}(r)$ does not vary too much with $\rho$ or $\rm{Pe}$ (see Figs.~\ref{fig:2} \& ~\ref{fig:4}), which likely explains the resemblance between Figs.~\ref{fig:5a} \& \ref{fig:6a}, and between Figs.~\ref{fig:5b} \& \ref{fig:6c}. We note that alternative flow-based approaches for defining chemical potentials for active systems obey the Gibbs--Duhem equation under local density approximations.\cite{hermann2019, paliwal2018}

\subsubsection{Total pressure}\label{subsubsec:TotalPressure}
We compare these \textcolor{black}{effective pressures} with the total pressure, $P$, of the active systems. Note that we substitute $k_{\rm{B}} T/\varepsilon=0.01$ into eq.~(\ref{eq:14}), as this was used in these LAMMPS simulations –-- this only affects the ideal gas pressure term, which we find to be small compared to the other terms, in the presence of significant activity.

In Figs.~\ref{fig:6b} \& \ref{fig:6d}, we plot the total pressure and its constituents: the swimming pressure and the passive pressure (the sum of the ideal and interaction terms). The general trend is for the swimming and (to a lesser extent) passive pressures (and hence the total pressure) to increase with increasing $\rho$ and $\rm{Pe}$. Consequently, at higher $\rho$ and $\rm{Pe}$ values, the total pressure is order of magnitudes larger than the \textcolor{black}{effective pressure}. It is expected from its coefficient that the swimming pressure should increase with $\rm{Pe}$ (Fig.~\ref{fig:6d}), in addition to $\dot{\textcolor{black}{\bm{r}}}_i$ also depending on $\textcolor{black}{F_{\rm{a}}}$ [eq.~(\ref{eq:9})]. The summation over the $N$ particles explains why the swimming pressure increases with $\rho$ (Fig.~\ref{fig:6b}). The passive pressure likely increases with density due to particles being forced closer together (Fig.~\ref{fig:4a}), such that more particles lie at distances with large $-\mathrm{d}u(r)/\mathrm{d}r$ values. For the WCA potential, the passive pressure increases with $\rm{Pe}$ (Fig.~\ref{fig:6d}). Correspondingly, it can be seen in Fig.~\ref{fig:2a} that the position of the minimum in $u_{\rm{eff}}(r)$ shifts to smaller $r$ values as $\rm{Pe}$ increases. The passive pressure for the shoulder potential simulations was similar at both $\rm{Pe}=120$ and $\rm{Pe}=300$.

\section*{Conclusions}\label{sec:Conclusions}
We have newly demonstrated that, for ABPs, a steady state $g(r)$ can be inverted to obtain $\beta u_{\rm{eff}}(r)$. This works for different potential types, P\'eclet numbers and densities tested. The $u_{\rm{eff}}(r)$ combines the underlying $u(r)$ and activity. The influence of these factors on $u_{\rm{eff}}(r)$ was demonstrated for a variety of potentials, with systematic changes in potential types, $\rm{Pe}$ and density. The inverse method converges quickly because, at low density, $u_{\rm{eff}}(r)$ resembles $\textcolor{black}{w_{\rm{eff}}(r)}$, which is used as the initial guess. Through $u_{\rm{eff}}(r)$, we see the non-additive effects of activity. The purely repulsive shoulder and WCA potentials give particularly insightful $u_{\rm{eff}}(r)$, since their $u_{\rm{eff}}(r)$ contain attractive wells: this is a clear demonstration of the emergent attractive behaviour of ABPs. Furthermore, the shape, such as the depth, of these attractive wells offers a means to quantify this emergent behaviour. Varying density, whilst keeping the $u(r)$ and activity parameters used in the LAMMPS simulation constant, produces density-dependent $u_{\rm{eff}}(r)$ that offer a way to quantify density-dependent emergent activity effects.

We therefore believe that this work presents another way to characterise and understand the behaviour of active particle suspensions –-- in particular, their structure. Whilst $u_{\rm{eff}}(r)$ could be viewed as a proxy for $g(r)$, this work raises the question, is there any more meaning to $u_{\rm{eff}}(r)$? Here, we have used $u_{\rm{eff}}(r)$ to calculate $\mu_{\rm{eff}}$ and $P_{\rm{eff}}$, which may help to address thermodynamic questions about active systems. Further work could explore if $u_{\rm{eff}}(r)$ could give any meaningful insight into the particle dynamics, and any quantitative information on the activity parameters, given the $u(r)$ used in the numerical simulation. It has been demonstrated with passive systems that the test-particle insertion inverse method can be extended to many-body interactions, i.e., finding $u^{n}(r)$, where $n$ denotes the $n$-body interaction.\cite{stones2023} This could also be explored with ABPs, finding $u^{n}_{\rm{eff}}(r)$.\cite{evans2024} We expect these higher-order interaction terms to be non-zero, since we have found \textcolor{black}{$u_{\rm{eff}}(r)$} to be density-dependent. 

\textcolor{black}{Another route for future exploration, currently underway, is using our inverse method with mixtures of active and passive particles. Effective interactions between active particles in active--passive mixtures have been previously found in the dilute limit,\cite{mu2022} whilst our method offers an approach applicable also beyond this limit, and to find effective interactions between all combinations of particle types. In addition, our method could be applied to other classes of active matter besides ABPs, such as run-and-tumble particles.\cite{santra2020} We expect the method to work for any active system that obtains steady-state structures and does not have areas of high local density. This includes systems with non-reciprocal interactions. It is worth mentioning that, in real systems, active particles may experience non-reciprocal interactions, either originating from hydrodynamic interactions\cite{bililign2022} or differences in surface mobility towards chemical gradients.\textcolor{black}{\cite{meredith2020}} Our test-particle insertion method ignores any potential non-reciprocity in the interactions; it just finds an effective (reciprocal) pair potential from the structure. We still expect our method to be able to find a $u_{\rm{eff}}(r)$, and thus also a $\mu_{\rm{eff}}$ and $P_{\rm{eff}}$. These might show how asymmetries in effective interactions shift collective steady states.}

\section*{Author contributions}
\textbf{Clare R. Rees-Zimmerman:} Conceptualization, Data curation, Formal analysis, Investigation, Methodology, Project administration, Software, Validation, Visualization, Writing – original draft, Writing – review \& editing. \textbf{C. Miguel Barriuso Gutierrez:} Data curation, Formal analysis, Investigation, Methodology, Project administration, Software, Validation, Writing – review \& editing. \textbf{Chantal Valeriani: }Conceptualization, Funding acquisition, Investigation, Methodology, Project administration, Resources, Supervision, Writing – review \& editing. \textbf{Dirk G. A. L. Aarts:} Conceptualization, Formal analysis, Funding acquisition, Investigation, Methodology, Project administration, Resources, Supervision, Validation, Writing – review \& editing.

\section*{Conflicts of interest}
There are no conflicts to declare.

\section*{Data availability}
The code for the test-particle insertion method can be found at https://github.com/creeszimmerman/TPI. 

\section*{Acknowledgements}
C. R. R. Z.’s work is funded by a Junior Research Fellowship from Christ Church, University of Oxford. The authors would like to thank Jos\'e Mart\'in-Roca (Universidad Complutense de Madrid) for checking the active pressure calculations. 
C. V. acknowledges \textcolor{black}{the funding awards} IHRC22/00002 and  PID2022-140407NB-C21 \textcolor{black}{from} MCIN/AEI /10.13039/501100011033 and FEDER, \textcolor{black}{EU}. \textcolor{black}{The authors thank Francisco Alarc\'on (University of Guanajuato) for useful discussions.}

\balance

\bibliography{rsc}
\bibliographystyle{rsc}

\end{document}


\pagestyle{fancy}
\thispagestyle{plain}
\fancypagestyle{plain}{
\renewcommand{\headrulewidth}{0pt}
}

\makeFNbottom
\makeatletter
\renewcommand\LARGE{\@setfontsize\LARGE{15pt}{17}}
\renewcommand\Large{\@setfontsize\Large{12pt}{14}}
\renewcommand\large{\@setfontsize\large{10pt}{12}}
\renewcommand\footnotesize{\@setfontsize\footnotesize{7pt}{10}}
\makeatother

\renewcommand{\thefootnote}{\fnsymbol{footnote}}
\renewcommand\footnoterule{\vspace*{1pt}%
\color{cream}\hrule width 3.5in height 0.4pt \color{black}\vspace*{5pt}} 
\setcounter{secnumdepth}{5}

\makeatletter 
\renewcommand\@biblabel[1]{#1}            
\renewcommand\@makefntext[1]%
{\noindent\makebox[0pt][r]{\@thefnmark\,}#1}
\makeatother 
\renewcommand{\figurename}{\small{Fig.}~}
\sectionfont{\sffamily\Large}
\subsectionfont{\normalsize}
\subsubsectionfont{\bf}
\setstretch{1.125}
\setlength{\skip\footins}{0.8cm}
\setlength{\footnotesep}{0.25cm}
\setlength{\jot}{10pt}
\titlespacing*{\section}{0pt}{4pt}{4pt}
\titlespacing*{\subsection}{0pt}{15pt}{1pt}

\fancyfoot{}
\fancyfoot[LO,RE]{\vspace{-7.1pt}}
\fancyfoot[CO]{\vspace{-7.1pt}\hspace{13.2cm} }
\fancyfoot[CE]{\vspace{-7.2pt}\hspace{-14.2cm} }
\fancyfoot[RO]{\footnotesize{\sffamily{1--\pageref{LastPage} ~\textbar  \hspace{2pt}\thepage}}}
\fancyfoot[LE]{\footnotesize{\sffamily{\thepage~\textbar\hspace{3.45cm} 1--\pageref{LastPage}}}}
\fancyhead{}
\renewcommand{\headrulewidth}{0pt} 
\renewcommand{\footrulewidth}{0pt}
\setlength{\arrayrulewidth}{1pt}
\setlength{\columnsep}{6.5mm}
\setlength\bibsep{1pt}

\makeatletter 
\newlength{\figrulesep} 
\setlength{\figrulesep}{0.5\textfloatsep} 

\newcommand{\topfigrule}{\vspace*{-1pt}%
\noindent{\color{cream}\rule[-\figrulesep]{\columnwidth}{1.5pt}} }

\newcommand{\botfigrule}{\vspace*{-2pt}%
\noindent{\color{cream}\rule[\figrulesep]{\columnwidth}{1.5pt}} }

\newcommand{\dblfigrule}{\vspace*{-1pt}%
\noindent{\color{cream}\rule[-\figrulesep]{\textwidth}{1.5pt}} }

\makeatother

\twocolumn[
  \begin{@twocolumnfalse}
{\hfill\raisebox{0pt}[0pt][0pt]{}\\[1ex]
 }\par
\vspace{1em}
\sffamily
\begin{tabular}{m{1.5cm} p{15cm} }

 &
\noindent\LARGE{\textbf{Supplementary information to `Effective interactions in active Brownian particles'}} \\
\vspace{0.3cm} & \vspace{0.3cm} \\

 & \noindent\large{Clare R. Rees-Zimmerman,$^{\ast}$\textit{$^{a}$} C. Miguel Barriuso Gutierrez,\textit{$^{b\ddag}$} Chantal Valeriani\textit{$^{b\ddag}$} and Dirk G. A. L. Aarts\textit{$^{a}$}} \\

\end{tabular}

 \end{@twocolumnfalse} \vspace{0.6cm}]

\renewcommand*\rmdefault{bch}\normalfont\upshape
\rmfamily
\section*{}
\vspace{-1cm}

\footnotetext{\textit{$^{a}$~Physical and Theoretical Chemistry Laboratory, University of Oxford, South Parks Road, Oxford OX1 3QZ, United Kingdom. Email: clare.rees-zimmerman@chch.ox.ac.uk or dirk.aarts@chem.ox.ac.uk}}
\footnotetext{\textit{$^{b}$~Departamento de Estructura de la Materia, Física Térmica y Electrónica, Universidad Complutense de Madrid, 28040 Madrid, Spain. Email: carbarri@ucm.es or cvaleriani@ucm.es}} 

\footnotetext{\ddag~Also at Grupo Interdisciplinar Sistemas Complejos, Madrid, Spain}

\appendix
\renewcommand\thefigure{\thesection.\arabic{figure}}

\section{Monte Carlo simulations}\label{appendix:MC}
The $\beta u_{\rm{eff}}(r)$ obtained for the active systems are used in Monte Carlo simulations of passive systems, to generate particle coordinates, and subsequently calculate the corresponding $g_{\rm{DH,MC}}(r)$. As a check, this is compared to the original $g_{\rm{DH}}(r)$ obtained from the LAMMPS simulations of the active particles to show that $\beta u_{\rm{eff}}(r)$ can be used to regenerate the structure.

Two-dimensional canonical simulations are carried out, fixing $N$, $V$ \& $T$, and hence allowing simulations at desired densities: this is helpful, since the density-dependence of $u_{\rm{eff}}(r)$ would be problematic for grand canonical simulations (Louis, 2002). The box size is chosen to house 2500 particles in a periodic box, as in the original LAMMPS simulations. Interactions up to a cut-off distance of \textcolor{black}{$r_{\rm{c}}=5\sigma$}, where $\sigma$ is the particle diameter, are included in calculations, with $r$ discretised into 500 equal-sized bins up to this distance.

Some trials ($10^6$) are run to obtain a first equilibrium snapshot, then a further equilibrium snapshot is saved every $10^4$ trials, until a total of $10^7$ trials have been run. In each trial move, the displacement of a random particle by a distance up to a maximum displacement, is either accepted or rejected. The maximum displacement is selected to obtain an acceptance rate in the range 25--50\%.\cite{frenkel2023}

\section{Inverse method convergence}\label{appendix:convergence}
\setcounter{figure}{0} 
Convergence checks were carried out for all parameter sets presented in the paper. As examples, convergence of the inverse method for the cases shown in Fig.~2 is demonstrated. Fig.~\ref{fig:B1a} shows excellent agreement between $g_{\rm{DH}}(r)$ and $g_{\rm{TPI}}(r)$, based on the converged $\beta u_{\rm{eff}}(r)$ found from the inverse method. As in passive particles examples,\cite{stones2019} Fig.~\ref{fig:B1b} shows convergence of $g_{\textcolor{black}{\rm{TPI,}}j}(r)$ to $g_{\rm{DH}}(r)$ to machine precision within 100 optimisations.

As described in Appendix~\ref{appendix:MC}, MC simulations are conducted using $\beta u_{\rm{eff}}(r)$, to verify that this can regenerate $g_{\rm{DH}}(r)$. A comparison of $g_{\rm{DH}}(r)$ and $g_{\rm{DH,MC}}(r)$ is shown in Fig.~\ref{fig:B1c}.  For each potential, a statistic measuring the difference between $g_{\rm{DH}}(r)$ and $g_{\rm{DH,MC}}(r)$, $\chi_{\rm{MC}}^2=\sum_{i}\left(g_{\rm{DH}} (r_i)-g_{\rm{DH,MC}} (r_i)\right) ^2$, is shown in Fig.~\ref{fig:B1c}. In each case, the difference is small.

\begin{figure}[h!]
\centering
   \begin{subfigure}[b]{0.45\textwidth}
     \subcaption{}
 \includegraphics[width=1\textwidth]{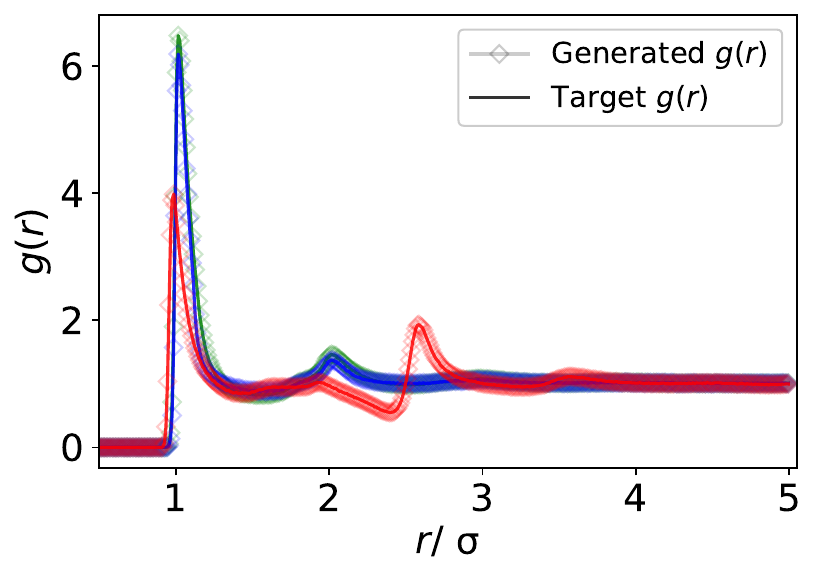}
        \label{fig:B1a}
    \end{subfigure}
    \quad
       \begin{subfigure}[b]{0.45\textwidth}
     \subcaption{}
    \includegraphics[width=1\textwidth]{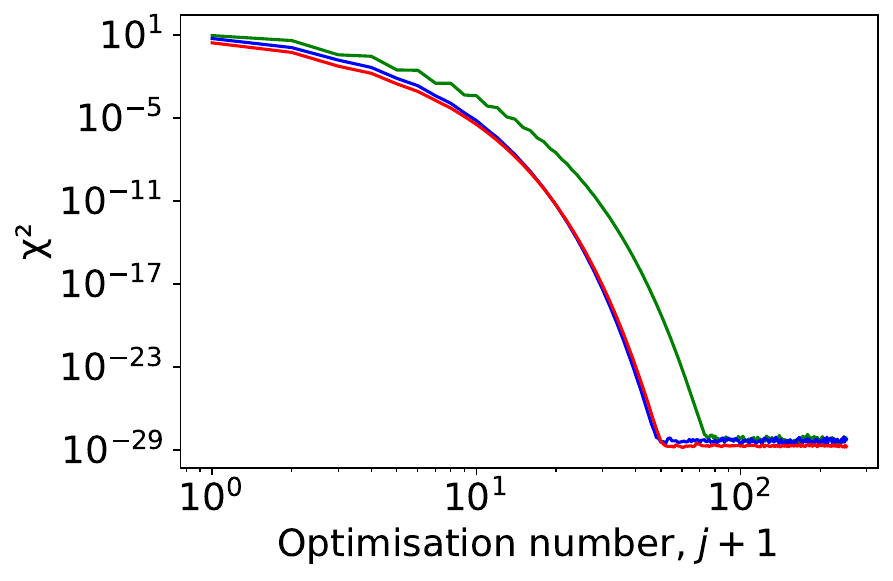}
        \label{fig:B1b}
    \end{subfigure}
      \quad
       \begin{subfigure}[b]{0.45\textwidth}
     \subcaption{}
     \includegraphics[width=1\textwidth]{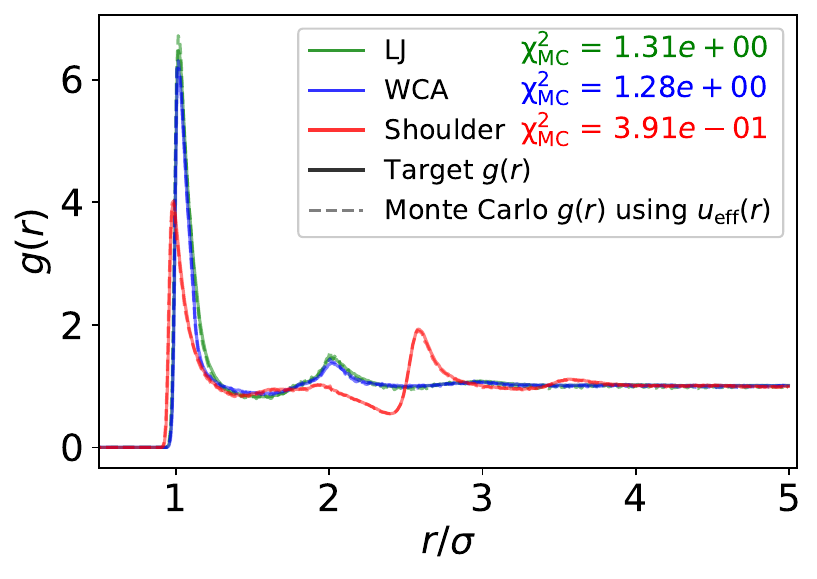}
        \label{fig:B1c}
    \end{subfigure}
  \caption{Demonstration of convergence for the systems shown in Fig.~2: LJ, WCA and shoulder potential at $\rho=0.3$ \& $\rm{Pe}=300$. (\subref{fig:B1a}) Plot comparing $g_{\rm{DH}}(r)$ and $g_{\rm{TPI}}(r)$, based on the converged $\beta u_{\rm{eff}}(r)$ from the inverse method. (\subref{fig:B1b}) The convergence in $g_{\textcolor{black}{\rm{TPI,}}j}(r_i)$ with optimisation number, $j+1$, for each potential. (\subref{fig:B1c}) Plot comparing $g_{\rm{DH}}(r)$ from the LAMMPS simulation, and $g_{\rm{DH,MC}}(r)$ based on a Monte Carlo simulation of $\beta u_{\rm{eff}}(r)$. The colour code in (\subref{fig:B1c}) also applies to subfigures (\subref{fig:B1a}) \& (\subref{fig:B1b}).}
  \label{fig:B1}
\end{figure}

\section{Comparison with the potential of mean force}\label{appendix:pmf}
\setcounter{figure}{0} 
A similar plot to Fig.~4 is presented, now for a passive LJ simulation, in Fig.~\ref{fig:C1}. This compares $u_{\rm{LJ}}(r)$ with \textcolor{black}{$w(r)$}. This is at $\rho=0.3$, exemplifying that, for passive systems, the potential of mean force explains much of the structure. Note that, in the passive LJ simulation here we use $k_{\rm{B}}T/\varepsilon=1$, to avoid phase separation,\cite{karakasidis2007} whilst the LAMMPS simulations in this work used $k_{\rm{B}} T/\varepsilon=0.01$, since these also included activity.

\begin{figure}[h!]
\centering
   \begin{subfigure}[b]{0.45\textwidth}
     \subcaption{}
    \hspace{-0.24cm}\includegraphics[width=1\textwidth]{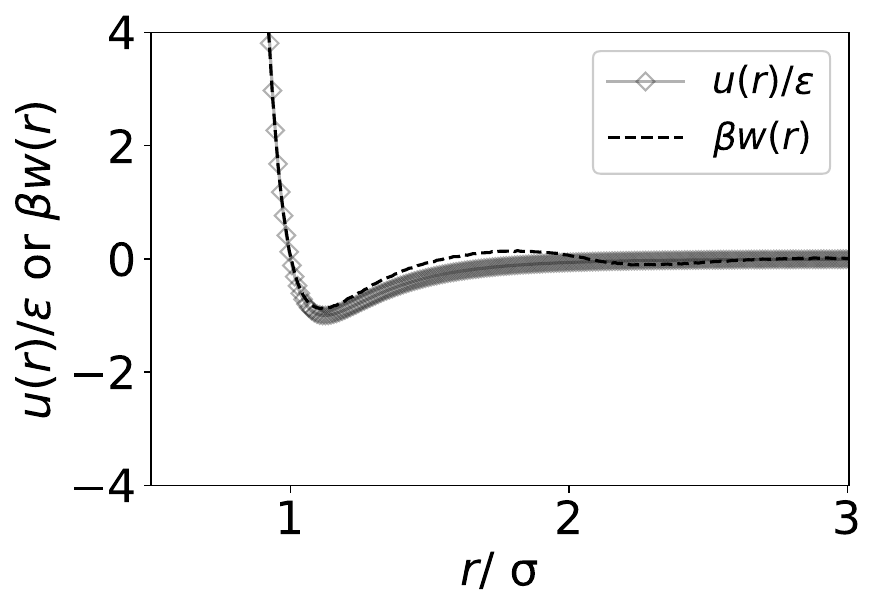}
        \label{fig:C1a}
    \end{subfigure}
    \quad
     \begin{subfigure}[b]{0.43\textwidth}
     \subcaption{}
\includegraphics[width=1\textwidth]{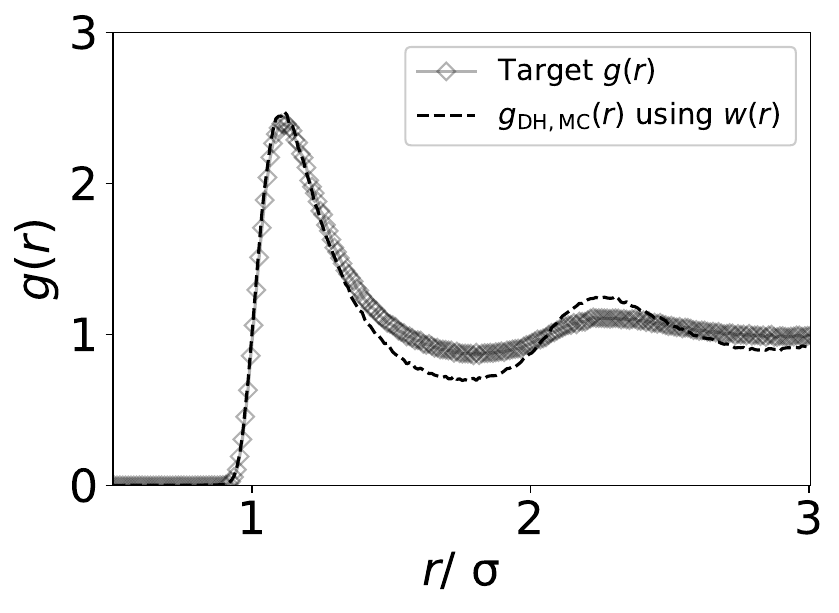}
        \label{fig:C1b}
    \end{subfigure}
  \caption{Demonstration of the closeness of $u_{\rm{LJ}}(r)$ with \textcolor{black}{$w(r)$}, using the example of a passive LJ simulation at $\rho=0.3$, $k_{\rm{B}} T/\varepsilon=1$: (\subref{fig:C1a}) Comparison of $u_{\rm{LJ}}(r)$ with \textcolor{black}{$w(r)$}. (\subref{fig:C1b}) Comparison of $g_{\rm{DH}}(r)$ from the original data with $g_{\rm{DH,MC}}(r)$ from a MC simulation using \textcolor{black}{$w(r)$}.}
  \label{fig:C1}
\end{figure}

\section{Comparison with the Gibbs--Duhem equation}\label{appendix:GD}
\setcounter{figure}{0} 
For the example shown in Fig.~5, with the shoulder potential and varying density, we attempt to compare with the Gibbs--Duhem equation at constant temperature, $\mathrm{d}\mu/\mathrm{d}\rho=(1/\rho)\mathrm{d}P/\rm{d}\rho$. A spline (‘scipy.interpolate.make\_smoothing\_spline’ in Python, with no regularisation parameter stated) is fitted to our data for $\mu_{\rm{ex}}(\rho)$, along with $\mu_{\rm{ex}}(\rho=0)=0$. This spline is differentiated to obtain an approximation for $\rm{d}\mu_{\rm{ex}}/\rm{d}\rho$. Using the trapezium rule from $\rho=10^{-7}$ to 0.4 in $2\times10^4$ steps, $\mathrm{d}P_{\rm{eff}}/\rm{d}\rho=\rho\rm{d}\mu_{\rm{ex}}/\rm{d}\rho$ is integrated numerically, assuming $P_{\rm{eff}}(\rho=0)=0$. In Fig.~\ref{fig:D1a}, we see that that $P_{\rm{eff,data}}$ and $P_{\rm{eff}}$ estimated using \textcolor{black}{the} Gibbs--Duhem equation agree well at low density ($\rho\le0.1$). As $\rho$ is increased, they increasingly disagree, though they do follow a similar general trend. However, the error in the estimate of $P_{\rm{eff}}$ also propagates with increasing $\rho$. As was discussed in the main text, the variation in $u_{\rm{eff}}(r)$ with $\rho$ may additionally explain the increasing disagreement as $\rho$ is increased.

Taking the equation for the free energy, $F=-PV+\mu N$,\cite{frenkel2023} we define an effective free energy for our 2D system of $F_{\rm{eff}}/A=-P_{\rm{eff}}+\mu_{\rm{eff}}\rho$. For interest, for this same data set, we plot $(F_{\rm{eff}}-\sigma^2 \mu^{\rm{o}} \rho)$ for a fixed area $A=\sigma^2$ as a function of $\rho$ in Fig.~\ref{fig:D1b}. The error bars in $P_{\rm{eff}}$ in Fig.~\ref{fig:D1a} are the same as those in Fig.~7a, whilst those in Fig.~\ref{fig:D1b} are derived from these. Thermodynamically, for a passive system, phase separation occurs when there is concavity in the free energy, $F$, i.e., a region where $\mathrm{d}^2 F/\rm{d}\rho^2<0$.\cite{barrat2003} Taking the error bars into account, Fig.~\ref{fig:D1b} does not particularly suggest concavity in $F_{\rm{eff}}(\rho)$ over the range of densities shown. Still, we stress that examining $F$ does not necessarily tell us about the phase behaviour of the ABPs. Furthermore, since $u_{\rm{eff}}(r)$ varies with $\rho$ (Fig.~5b), we cannot say if we can represent different densities with the same equivalent passive system without first examining many-body interactions --- so we also cannot state whether this could be a plot showing the stability of an equivalent passive system.

\begin{figure}[h!]
\centering
      \begin{subfigure}[b]{0.45\textwidth}
     \subcaption{}
    \includegraphics[width=1\textwidth]{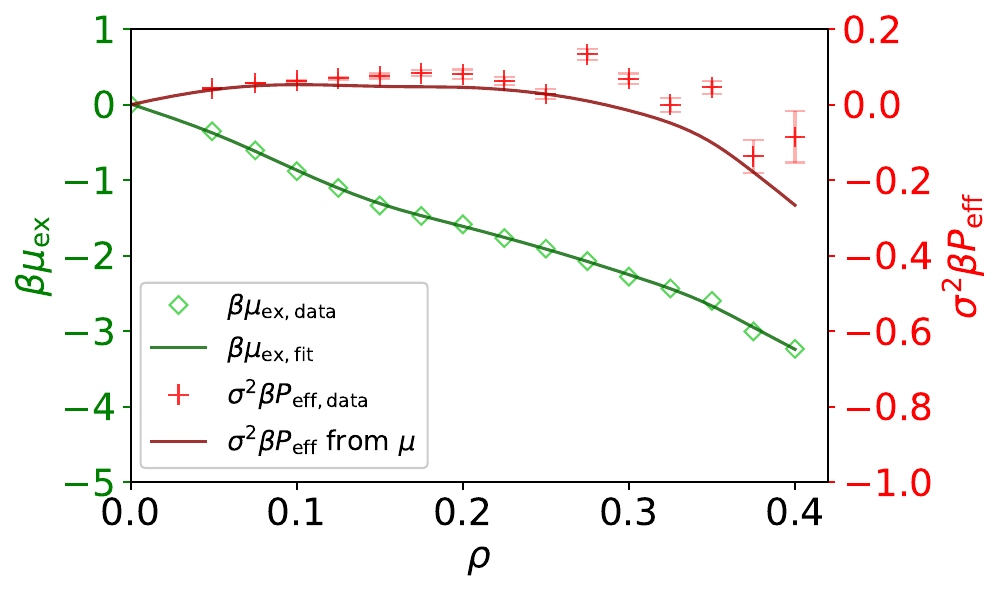}
        \label{fig:D1a}
    \end{subfigure}
      \quad
          \begin{subfigure}[b]{0.4\textwidth}
     \subcaption{}
\hspace{-0.85cm} \includegraphics[width=1\textwidth]{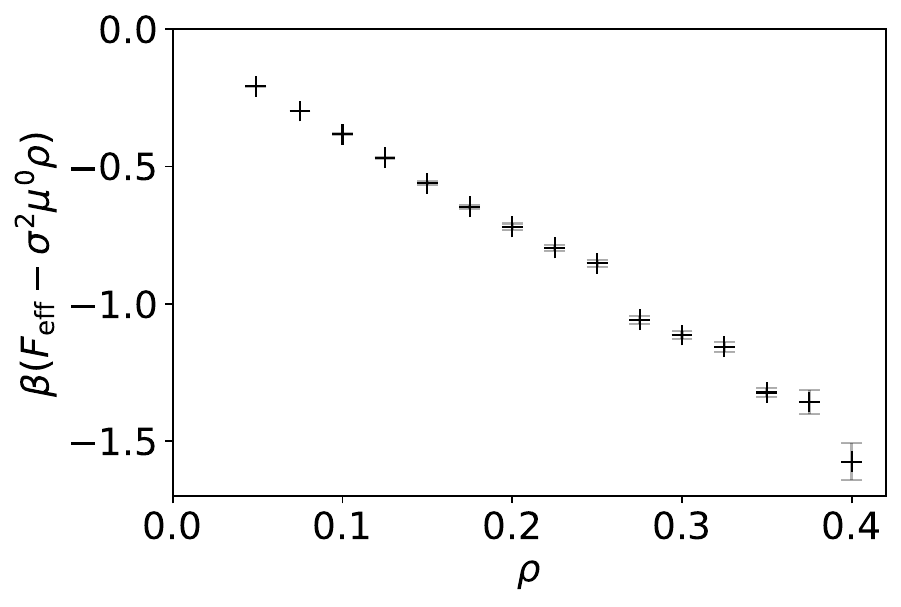}
        \label{fig:D1b}
    \end{subfigure}
  \caption{For ABPs with a shoulder potential at $\rm{Pe}=120$ and varying density, as in Fig.~5: (\subref{fig:D1a}) Plot of $\mu_{\rm{ex}}$ and $P_{\rm{eff}}$ as a function of $\rho$, comparing data values for $P_{\rm{eff}}$ to those estimated from the Gibbs--Duhem equation. (\subref{fig:D1b}) Plot of $F_{\rm{eff}}$ (for fixed area) as a function of $\rho$.}
  \label{fig:D1}
\end{figure}

We can similarly write the Gibbs--Duhem equation in terms of variation in P\'eclet number, $\mathrm{d}\mu/\mathrm{d}\mathrm{Pe}=(1/\rho)\mathrm{d}P/\mathrm{d}\mathrm{Pe}$, and consider if it holds for $\mu_{\rm{eff}}$ and $P_{\rm{eff}}$. We use the example of WCA potential at varying $\rm{Pe}$ (Fig.~3). The procedure is analogous to that conducted for Fig.~\ref{fig:D1a}. A spline (same type) is fitted to our data for $\mu_{\rm{ex}} (\rm{Pe})$. Differentiating the spline obtains an approximation for $\rm{d}\mu_{\rm{ex}}/\rm{d}\rm{Pe}$. Using the trapezium rule from $\rm{Pe}=6$ to 300 in $10^4$ steps, $\mathrm{d}P_{\mathrm{eff}}/\mathrm{d}\mathrm{Pe}=\rho \mathrm{d}\mu_{\rm{ex}}/\mathrm{d}\rm{Pe}$ is integrated numerically, beginning from an arbitrary value. We shift all predicted $P_{\rm{eff}}$ values to minimise the sum of the squared differences between the predicted and calculated $P_{\rm{eff}}$ values. We did this instead of picking a data point from which to integrate, due to the significant error bars in the calculated $P_{\rm{eff}}$ values. We also note that, due to the small value of $k_{\rm{B}}T/\varepsilon = 0.01$, the inversion at $\rm{Pe=0}$ is not completely accurate.\cite{reeszimmerman2025}

Fig.~\ref{fig:D1b} shows the resulting comparison of $P_{\rm{eff,data}}$ and $P_{\rm{eff}}$ estimated using Gibbs--Duhem equation. Within the error bars, there is reasonably good agreement. This is a really interesting result. However, it is difficult to interpret, as there is not an equivalent to the P\'eclet number for a passive system. We observe in Fig.~3a that $u_{\rm{eff}}(r)$ varies somewhat with Pe but, again, the relevance of this to the Gibbs--Duhem equation is hard to interpret.

\begin{figure}[h!]
\centering
  \includegraphics[width=0.45\textwidth]{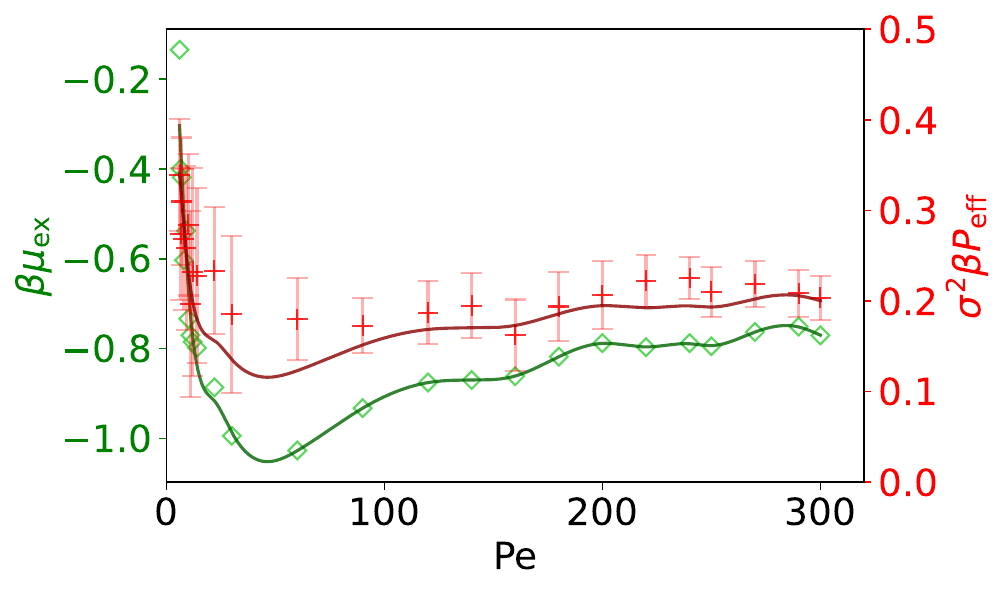}    
  \caption{Plot of $\mu_{\rm{ex}}$ and $P_{\rm{eff}}$ as a function of $\rm{Pe}$, comparing data values for $P_{\rm{eff}}$ to those estimated from the Gibbs--Duhem equation. This is for ABPs with a WCA potential at $\rho=0.3$ and varying $\rm{Pe}$, as in Fig.~3. The key is the same as that in Fig.~\ref{fig:D1a}.}
  \label{fig:D2}
\end{figure}

\section{Second virial coefficient}\label{appendix:virial}
\setcounter{figure}{0} 
\textcolor{black}{It is interesting to consider whether the second virial coefficient $B_2$ might provide any information about the phase behaviour of our ABPs. For our 2D systems, an effective $B_2$ (based on $u_{\mathrm{eff}}(r)$) is calculated using}
\begin{equation}
\textcolor{black}{B_2=-\frac{1}{2}\int_{0}^{\infty} \left(\exp[-\beta u_{\mathrm{eff}}(r)]-1 \right) 2 \pi r\mathrm{d}r,}
\end{equation}
\textcolor{black}{which we evaluate numerically using Simpson's rule from $r=0$ up to our inversion cutoff of \textcolor{black}{$r_{\rm{c}}=5\sigma$.}}
\textcolor{black}{By generalising the Vliegenthart--Lekkerkerker (VL) relation, criticality is expected for LJ-like potentials around}
\begin{equation}
\textcolor{black}{B_2/\nu_0\approx-8.2,}
\end{equation}
\textcolor{black}{where $\nu_0=\pi \sigma^2/4$.\textcolor{black}{\cite{turci2021}}}

\begin{figure}[h!]
     \centering
     \begin{subfigure}[b]{0.45\textwidth}
         \centering
         \subcaption{}
         \includegraphics[width=\textwidth]{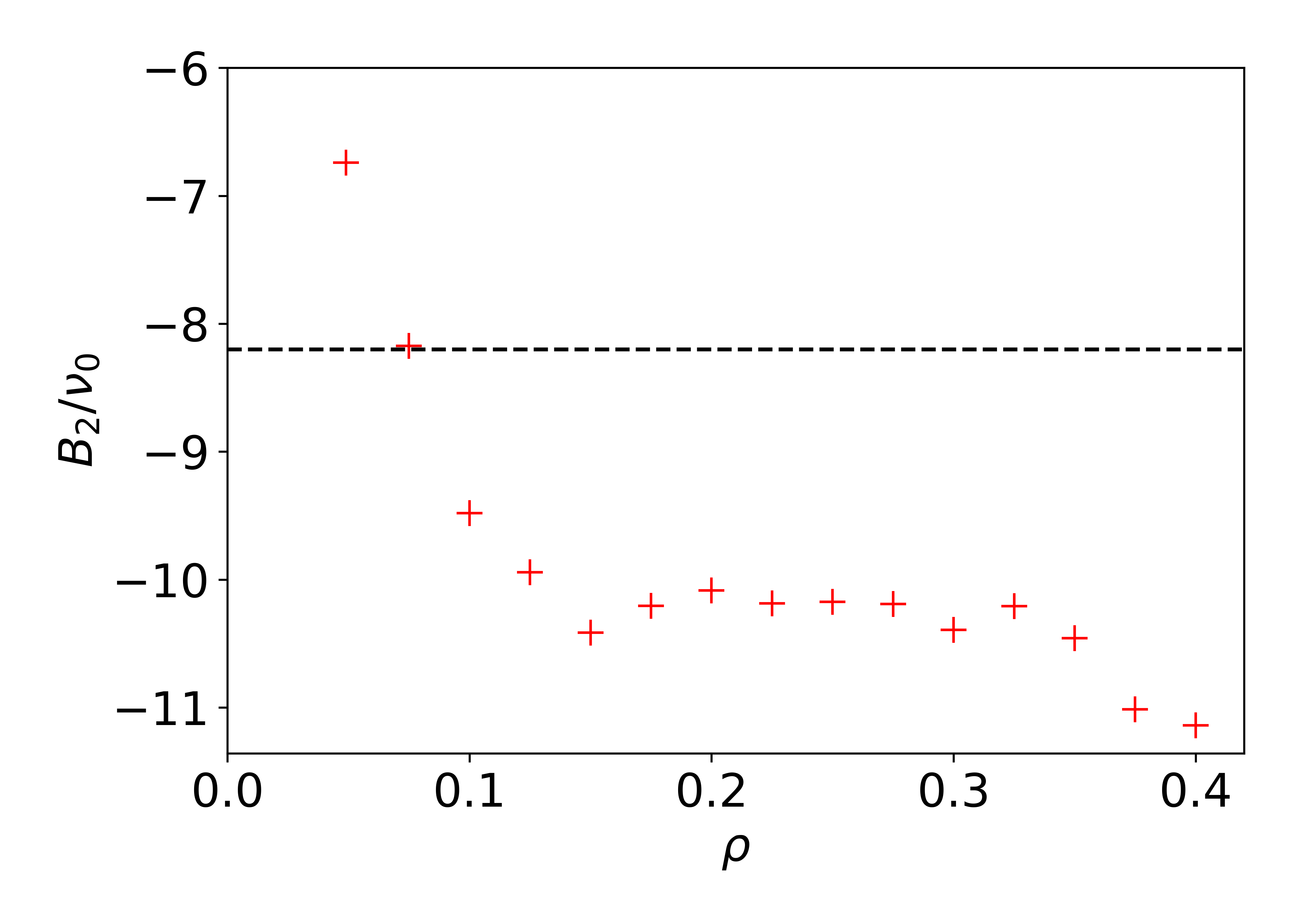}
         \label{fig:B2_b}
     \end{subfigure}
     \begin{subfigure}[b]{0.45\textwidth}
         \centering
         \subcaption{}
         \includegraphics[width=\textwidth]{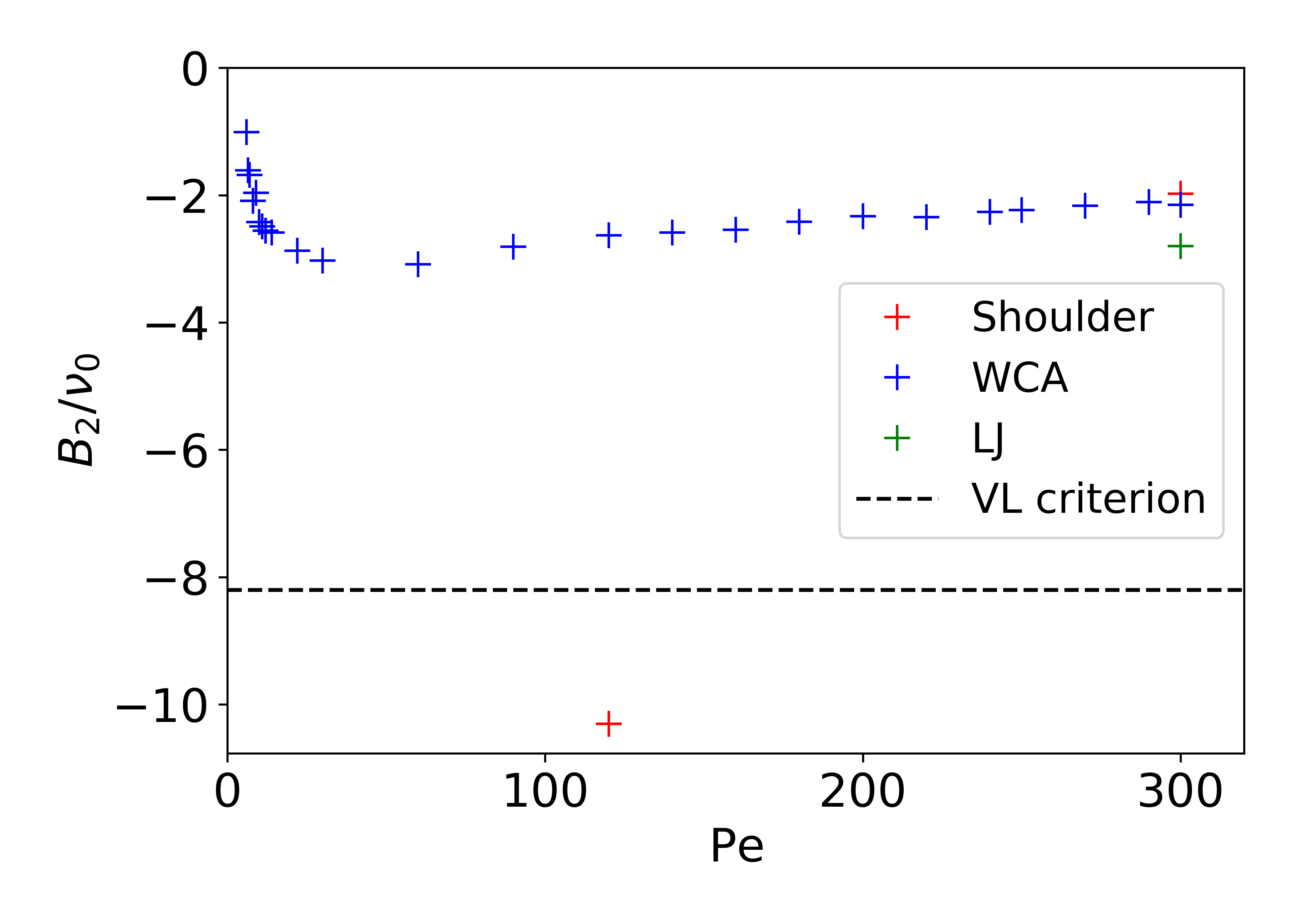}
         \label{fig:B2_a}
     \end{subfigure}
\caption{\textcolor{black}{Plot of effective $B_2/\nu_0$ as a function of (\subref{fig:B2_b}) $\rho$ for the shoulder potential at $\rm{Pe}=120$ (as in Fig.~5) and (\subref{fig:B2_a}) $\rm{Pe}$ for the shoulder, WCA and LJ potentials at $\rho=0.3$. An approximate critical value of $B_2/\nu_0\approx-8.2$ is shown by a dashed black line.}}
    \label{fig:B2_virial}
\end{figure}

\textcolor{black}{We note similarity in general graph shapes between Fig.~\ref{fig:B2_virial} (effective $B_2$) and Figs.~6 (effective $\mu$),~7a and 7c (effective $P$) in the main text. This is more so in the case of the plots against $\rho$ than those against $\rm{Pe}$: the initial decrease in the effective $B_2$ with $\rho$ is not seen in the effective $\mu$ or $P$.}

\textcolor{black}{We note that all the LJ and WCA simulations run have an effective $B_2$ less negative than the approximate critical value (Fig.~\ref{fig:B2_a}). This matches with our snapshots for WCA and LJ appearing to show a single phase (see Appendix~\ref{appendix:snapshots}).}

\textcolor{black}{At $\rm{Pe=120}$, the effective $B_2$ values for the shoulder potential simulations are less negative than this approximate critical value only for low density (Fig.~\ref{fig:B2_b}), whilst at $\rho=0.3$, the effective $B_2$ is less negative than the approximate critical value only at high $\rm{Pe}$ ($\rm{Pe=300}$, Fig.~\ref{fig:B2_a}). This may at first seem at odds with our snapshots of the shoulder potential, which seem to show single phase across all densities and $\rm{Pe}$ run. Nevertheless, we note that this approximate critical value was found for LJ-like potentials, which does not describe the shoulder potential well. Thus, we conclude that this LJ-based critical value need not apply quantitatively for the shoulder potential; a more negative critical $B_2$ value might be required, should phase separation occur.}

\textcolor{black}{\citet{turci2021} studied 2D (Supplementary Information) and 3D (main text) ABP systems. They found effective pair potentials but of a fundamentally different definition: we find a state-dependent $u_{\rm{eff}}(r)$ that alone exactly reproduces $g(r)$ at a given state point, whilst \citet{turci2021} find exact 2,3,...$N$-body effective potentials that together reproduce the structure of the ABP system. Our $B_2$ calculations are therefore also different; ours are based on $u_{\rm{eff}}(r)$, whereas theirs uses their effective 2-body interaction.}

\textcolor{black}{\citet{turci2021} found their effective 2-body potential to by itself not predict phase separation (resulting in $B_2$ values less negative than the approximate critical value), even when the original ABP snapshots did show this. They concluded that many-body effective interactions must therefore be needed to account for the phase separation; their effective 2-body potentials alone were insufficiently attractive. Since test-particle insertion struggles with areas of high density, we could not study phase-separated examples, so we cannot say whether our work agrees with this conclusion of theirs. We also cannot approach MIPS from the dilute side without encountering high local densities. We hypothesise that, since our $u_{\rm{eff}}(r)$ does account for all effective higher-order interactions, that it likely would predict the correct phase behaviour (reproducing $g(r)$) at the given state point. However, a $u_{\rm{eff}}(r)$ derived for a given density might not accurately predict the phase behaviour at a different density.}

\textcolor{black}{Another thought is that finding an orientation-dependent $u_{\rm{eff}}(r,\theta)$, where $\theta$ is the angle between the self-propulsion force of two particles, might better describe the physics of the ABPs, and thus potentially better predict MIPS. \citet{miranda2025} (using an expression for the low-density limit) found $u_{\rm{eff}}(r,\theta)$ to only be attractive in a certain angular region, around the front of the particle. Extending test-particle insertion to anisotropic interactions is currently work-in-progress.}

\section{Mean free path}\label{appendix:freepath}
\setcounter{figure}{0} 
\textcolor{black}{Since free active particles do not move in straight lines, we  first show the mean free time between collisions, $\overline{\tau}_{\rm{f}}$ (Figs.~\ref{fig:F1}a and c), as opposed to the mean free path. We define a collision between two particles as 
$r_{ij} < 2^{1/6}\sigma$. For WCA interactions, this exactly marks the cutoff for repulsion. For LJ interactions, this criterion corresponds with distances smaller than the potential minimum at $r_{ij} = 2^{1/6}\sigma$, providing a practical threshold for entering the strongly repulsive regime and ensuring that only significant overlaps are counted as collisions. For the shoulder potential, we define a collision using the same criterion as for the LJ and WCA systems, $r_{ij} < 2^{1/6}\sigma$, where
$\sigma$ is the core diameter, maintaining consistency with the LJ/WCA analysis and allowing a uniform calculation of mean free paths across different potentials. The longer-range shoulder at $\sigma_{\rm{s}}$ is not included in the collision definition. }

\setcounter{figure}{0} 
\begin{figure*}[h!]
    \centering
    \includegraphics[width=\textwidth]{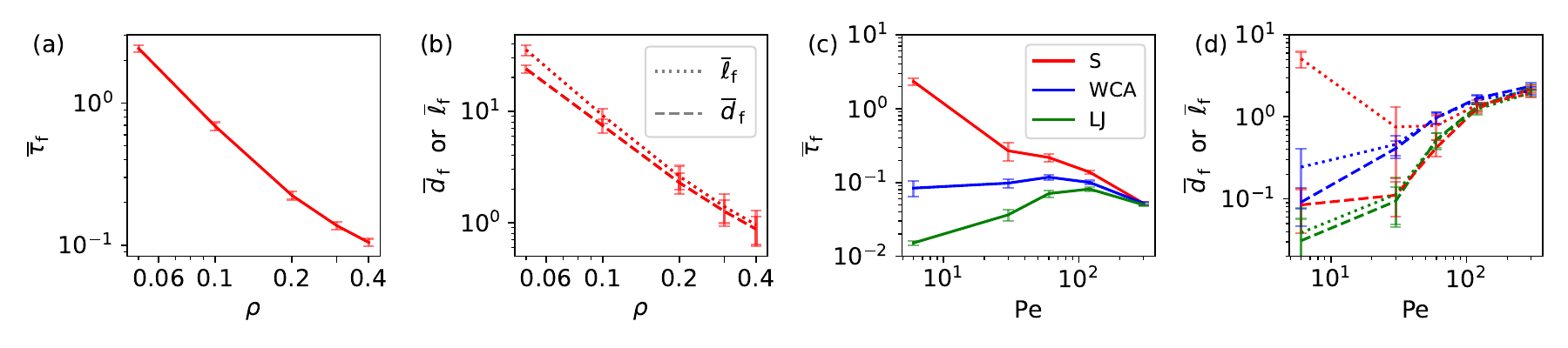}
    \caption{\textcolor{black}{Plot of the mean free time, $\overline{\tau}_{\rm{f}}$, as a function of (a) the density, $\rho$, for the shoulder simulations (at fixed $\rm{Pe}=120$), and (c) the P\'eclet number, $\rm{Pe}$, for the shoulder, WCA and LJ simulations (at fixed $\rho=0.3)$. In addition, we plot the mean free distance, $\overline{d}_{\rm{f}}$, and mean free path length, $\overline{\ell}_{\rm{f}}$, as a function of (b) density and (d) the P\'eclet number. The error bars show the standard deviation of the distribution of free times between collisions. Error bars are presented scaled by a factor of $0.05$ for (a), (b), and (c) and by a factor of $0.025$ for (d). The collision condition is set to $r_{ij}<2^{1/6}\sigma$. The keys in (b) and (c) also apply to (d).}}
    \label{fig:F1}
\end{figure*}

\textcolor{black}{For the shoulder potential simulations at varying density, $\overline{\tau}_{\rm{f}}$ decreases as the density is increased (Fig.~\ref{fig:F1}a), as expected. We observe non-monotonic behaviour in $\overline{\tau}_{\rm{f}}$ when increasing $\rm{Pe}$ for the WCA and LJ potentials (Fig.~\ref{fig:F1}c). For the WCA potential, at low $\rm{Pe}$, $\overline{\tau}_{\rm{f}}$ is low due to the fairly homogeneous distribution of particles. The LJ potential has a low $\overline{\tau}_{\rm{f}}$ at low $\rm{Pe}$ for a different reason: our cutoff for counting collisions falls at the minimum of the LJ potential. Collisions are therefore counted to occur very frequently, as particles oscillate around the bottom of this well. For both WCA and LJ, as $\rm{Pe}$ is first increased, clusters grow in size, leaving more free space for particles to travel. However, at higher $\rm{Pe}$ still, $\overline{\tau}_{\rm{f}}$ then decreases because particles move faster due to their higher activity, and clusters break. The snapshots in Fig~\ref{fig:E1} subtly show this. For the shoulder potential at low $\rm{Pe}$, the structures are initially `frustrated', with a sharp, narrow distribution of interparticle distances (just past the soft shell diameter). As $\rm{Pe}$ is increased, the activity enables more particles to penetrate the soft shell (at $r=2.5\sigma$), and so $\overline{\tau}_{\rm{f}}$ monotonically decreases.}

\textcolor{black}{Additionally, we compare the mean free path distance $\overline{d}_{\rm{f}}$ with the mean free path length $\overline{\ell}_{\rm{f}}$ (Fig.~\ref{fig:F1}b and d). Note that, at each given density, $\overline{\ell}_{\rm{f}} \geq \overline{d}_{\rm{f}}$, with equality achieved in the high-$\rm{Pe}$ limit. Generally, the greater the relative importance of rotational diffusion to activity is (i.e., the lower $\rm{Pe}$ is), the greater the difference between $\overline{\ell}_{\rm{f}}$ and $\overline{d}_{\rm{f}}$ is. Furthermore, $\overline{\ell}_{\rm{f}}$ and $\overline{d}_{\rm{f}}$ both generally increase with $\rm{Pe}$, as particles move faster and with more persistence. At the highest $\rm{Pe}$ studied, activity dominates over the interaction potential, making $\overline{\tau}_{\rm{f}}$, $\overline{d}_{\rm{f}}$ and $\overline{\ell}_{\rm{f}}$ nearly independent of the interaction potential (Figs. \ref{fig:F1}c and d).} 

\textcolor{black}{An exception occurs for the shoulder potential. We observe that, at low $\rm{Pe}$, the difference between $\overline{\ell}_{\rm{f}}$ and $\overline{d}_{\rm{f}}$ is most noticeable for the shoulder potential. This is due to the frustrated structures: the shoulder of the potential causes weak deflections, without meeting the chosen collision criterion. The degree of frustration in the structures decreases as $\rm{Pe}$ is increased, leading to the observed non-monotonicity in $\overline{\ell}_{\rm{f}}$. Also, for the shoulder potential, as density is increased (at fixed $\rm{Pe=120}$), particles become more restricted and collide more frequently, leading to $\overline{\ell}_{\rm{f}}$ approaching $\overline{d}_{\rm{f}}$ (Fig.~\ref{fig:F1}b). }

\section{Snapshots of simulations}\label{appendix:snapshots}
\setcounter{figure}{0} 
\begin{figure*}[h!]
\centering
  \includegraphics[width=\textwidth]{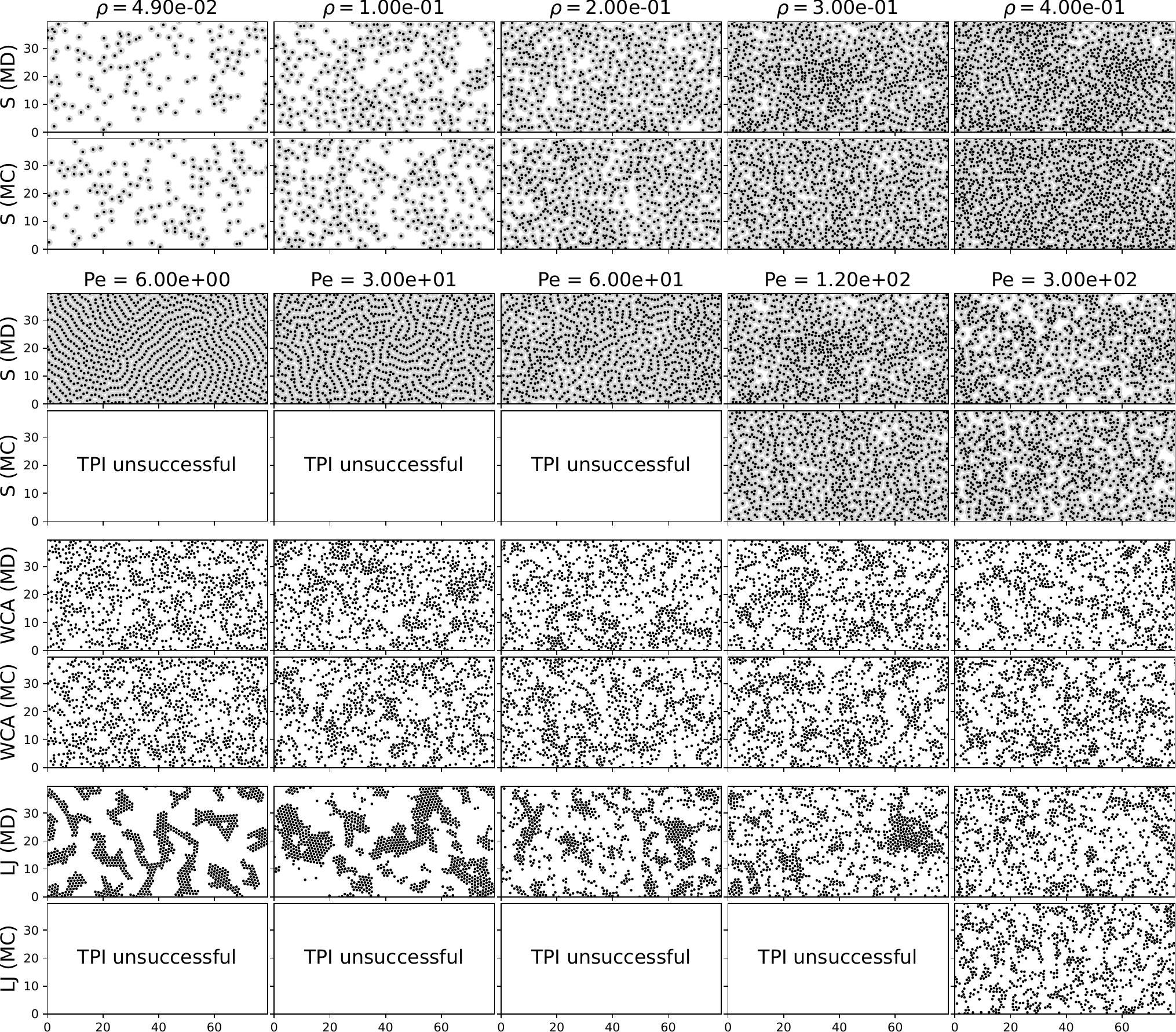}    
  \caption{\textcolor{black}{Example snapshots of (top half of each panel) the original MD simulations and (bottom half of each panel) MC simulations using $\beta u_{\rm{eff}}(r)$. The top row shows results for the shoulder potential at varying $\rho$ (at fixed $\rm{Pe}=120$), and the three bottom rows show the shoulder, WCA and LJ potentials at varying $\rm{Pe}$ (at fixed $\rho=0.3$). Note that only a cropped area of each snapshot is shown, for ease of comparison. This is a fuller version of  Fig.~1 in the main text.}}
  \label{fig:E1}
\end{figure*}
\textcolor{black}{Fig.~\ref{fig:E1} presents snapshots of particle coordinates produced via the original MD simulations, and also via MC simulations using $\beta u_{\rm{eff}}(r)$. Comparing these snapshots is highly interesting: the MD snapshots are obtained by simulating active particles, whilst the MC simulations are of equivalent passive systems (those which have the same $g(r)$).}

\textcolor{black}{Note that the areas of high local density for LJ at $\rm{Pe}<300$ meant that the inversion to $\beta u_{\rm{eff}}(r)$ was not successful (see Sec.~3.1.2 in the main text). Therefore, no MC snapshots are shown for these. Also, in the MD snapshots for the shoulder potential, there is local high density along the chains at $\rm{Pe}<120$, again resulting in unsuccessful inversion.}

\textcolor{black}{Visually, for the other cases when the inversion to $\beta u_{\rm{eff}}(r)$ was successful, there is generally excellent agreement between the structures from the original MD simulations, and those regenerated by MC simulation. This is expected, given the good agreement between $g_{\rm{DH}}(r)$ and $g_{\rm{DH,MC}}(r)$ (see Fig.~\ref{fig:B1c}). For all examples shown, $\chi^2_{\rm{MC}}<10$. For cases where $\chi^{2}_{\rm{MC}}\gtrsim 2.5$, slight differences between the MD and MC snapshots might become perceptible. For example, for the shoulder potential at $\rho=0.3$ and $\rm{Pe}=120$, remnant of the frustrated structures found at lower $\rm{Pe}$, TPI does not quite perfectly reproduce $g_{\rm{DH}}(r)$, leading to slight visual disagreement in the snapshots for this case. This is perhaps also noticeable for the shoulder potential at $\rho=0.4$ and $\rm{Pe}=120$. For the shoulder potential, agreement improves as $\rm{Pe}$ is increased further to $\rm{Pe}=300$. Note also that whilst our inverse method works by matching the pair correlations, it does not necessarily match any higher-order structural correlations, $g^n(r),\ n>2$. Nevertheless, this would not be detectable by eye in comparing these snapshots.}

\bibliography{rsc}
\bibliographystyle{rsc}